\documentstyle[11pt,aaspp4]{article}

\lefthead{Soo-Chang Rey et al.}
\righthead{RR Lyrae stars in the $\omega$ Centauri}

\begin{document}
\title{CCD PHOTOMETRY OF THE GLOBULAR CLUSTER\\
     $\omega$ CENTAURI.\\
       I. METALLICITY OF RR LYRAE STARS FROM $Caby$ PHOTOMETRY\\} 

\author{Soo-Chang Rey\altaffilmark{1}, Young-Wook Lee, and Jong-Myung Joo\altaffilmark{1}}
\affil{Center for Space Astrophysics \& Department of Astronomy,\\
    Yonsei University, Shinchon 134, Seoul 120-749, Korea\\
        Electronic mail : (screy, ywlee, jmjoo)@csa.yonsei.ac.kr}

\author{Alistair Walker}
\affil{Cerro Tololo Inter-American Observatory, NOAO, Casilla 603, La Serena, Chile\\
        Electronic mail : awalker@noao.edu}

\and

\author{Scott Baird}
\affil{Department of Physics \& Astronomy,\\
       Benedictine College, Atchison, Kansas 66002-1499,\\ 
       and Department of Physics \& Astronomy,\\
       University of Kansas, Lawrence, Kansas 66045-2151, USA\\
        Electronic mail : baird@kuphsx.phsx.ukans.edu}

\altaffiltext{1}{Visiting Astronomer, Cerro Tololo Inter-American Observatory,
National Optical Astronomy Observatories, which is operated by the Association
of Universities for Research in Astronomy, Inc. (AURA) under cooperative agreement
with the National Science Foundation.}

\begin{abstract}
We present new measurements of the metallicity of 131 RR Lyrae stars in the 
globular cluster $\omega$ Centauri, using the $hk$ index of the $Caby$ photometric 
system. The $hk$ method has distinct advantages over $\Delta$$S$ and other
techniques in determining the metallicity of RR Lyrae stars, and has allowed us
to obtain the most complete and homogeneous metallicity data to date for the
RR Lyrae stars in this cluster. For RR Lyrae stars in common with the
$\Delta$$S$ observations of Butler et al. (1978) and Gratton et al. (1986),
we have found that our metallicities, [Fe/H]$_{hk}$, deviate systematically from
their $\Delta$$S$ metallicity, while our [Fe/H]$_{hk}$ for
well observed field RR$ab$ stars are consistent with previous spectroscopic
measurements. We conclude that this is due to the larger errors associated with 
the previous $\Delta$$S$ observations for this cluster. The M$_{V}$(RR) - [Fe/H]
and period-shift - [Fe/H] relations obtained from our new data are consistent 
with the evolutionary models predicted by Lee (1991), confirming that the 
luminosity of RR Lyrae stars depends on evolutionary status as well as metallicity.
Using the period - amplitude diagram, we have also identified highly evolved
RR$ab$ stars in the range of -1.9 $\le$ [Fe/H] $<$ -1.5, as predicted from
the synthetic horizontal-branch models.
\end{abstract}


\section{INTRODUCTION}

For dating globular clusters and several other important problems
(e.g., measuring distances to Population II objects), it is essential to
know the luminosity of the RR Lyrae stars, M$_{bol}$(RR), and how it varies with
metal abundance (see Sandage 1990b; Lee, Demarque, and Zinn 1990, hereafter LDZ).
The variation of M$_{bol}$(RR) with [Fe/H] affects the age - metallicity
relation of the Galactic globular cluster system, and thus provides constraints
on the scenarios of the Galaxy formation. However, due to the variety of
different techniques used, the particular data set chosen, and the reddening 
corrections adopted, there is no consensus on the size 
of the dependency of M$_{bol}$(RR) upon [Fe/H] (Layden et al. 1996).
To investigate and resolve the problem of the dependence of M$_{bol}$(RR) 
on [Fe/H], one needs a large sample of RR Lyrae stars, spanning a wide range
of [Fe/H], for which precise measurements of relative luminosity and [Fe/H] 
exist. The RR Lyrae stars in $\omega$ Cen are an ideal sample for
this study. In $\omega$ Cen, there is a wide range in [Fe/H], and clearly the
relative values of M$_{bol}$(RR) can be inferred straightforwardly from their
mean apparent visual magnitudes since they are all located at the same
distance and are all reddened by the same amount.

However, investigations by Freeman \& Rodgers(1975), Butler et al. (1978, hereafter BDE),
Sandage (1982), and Gratton et al. (1986, hereafter GTO) have revealed that
the M$_{bol}$(RR) - [Fe/H] correlation in $\omega$ Cen is peculiar: 
a few metal-rich $({\rm[Fe/H]} >  -1.1)$ RR Lyrae stars in their sample
are fainter than the more metal-poor ones, but no obvious correlation
exists among the metal-poor $({\rm[Fe/H}] < -1.4)$ RR Lyrae stars.
This and the lack of a period-shift - [Fe/H] correlation amongst the
variables was recognized by Sandage (1982) as a possible contradiction 
to his steep correlation between M$_{bol}$(RR) and [Fe/H].  In general
these ``discrepant'' observational results were simply considered to be yet another
anomaly of the stellar population of $\omega$ Cen (see also Smith 1995).

Recent advances in our understanding of the evolution of horizontal-branch (HB)
stars are throwing new light on this long-standing problem. In particular, the
HB evolutionary models by Lee (1990) suggest that M$_{bol}$(RR) depends
on HB morphology as well as metallicity, especially when the HB
morphology is extremely blue, due to the effect of redward evolution off
the zero-age horizontal-branch (ZAHB).  Using these model calculations, 
Lee (1991) has shown that the observed nonlinear behavior of M$_{bol}$(RR)
with [Fe/H] in $\omega$ Cen is not something peculiar, but is in fact
predicted.  The detailed model calculations suggest that two effects are
responsible for the observed behavior of M$_{bol}$(RR) with [Fe/H] in 
$\omega$ Cen, and are:  (1) the abrupt increase in M$_{bol}$(RR) near 
[Fe/H] = -1.5 as RR Lyrae stars 
become highly evolved stars from the blue side of the instability strip as HB
morphology gets bluer with decreasing [Fe/H], and (2) the nonmonotonic
behavior of the HB morphology with decreasing [Fe/H], which together with
the first effect makes the correlation between M$_{bol}$(RR) and [Fe/H] looks
like a step function, because M$_{bol}$(RR) depends sensitively on HB morphology.
Despite the lack of a complete understanding of why HB morphology changes as
it does, the definite conclusions from Lee's (1991) work are:
(1) The correlation between M$_{bol}$(RR) and [Fe/H] in the halo of our Galaxy
is probably not linear due to the effect of HB morphology (evolution).
(2) The use of a simple linear relationship between M$_{bol}$(RR) and
[Fe/H] in deriving the distances to blue HB clusters should be avoided.
This suggests that when the distances to the population II objects are to be 
estimated using the RR Lyrae stars, the HB type of the stellar population, 
as well as metallicity, must be known.

Although the $\omega$ Cen data do appear to support this model, a definite 
conclusion was not possible because of the uncertainty in [Fe/H] and of the lack of
metal-rich stars in the available data. In order to provide a more complete and 
homogeneous sample of RR Lyrae stars with
relatively well-measured metallicity, we obtained [Fe/H] abundances for most of the
RR Lyrae stars in $\omega$ Cen, using the $Caby$ photometric system.
The $Caby$ photometric system is an expansion of the standard $uvby$ system 
with the inclusion of a fifth filter, $Ca$, centered on the K and H lines of Ca II
(90 $\AA$ FWHM).
The $hk$ index is defined as $hk$ =
$(Ca-b)-(b-y)$, and is found to be much more sensitive to metal abundance than
the $Str$$\ddot o$$mgren$ $m_{1}$ index (Anthony-Twarog et al. 1991; Twarog \& 
Anthony-Twarog 1991, 1995; Anthony-Twarog \& Twarog 1998).
The sensitivity of the $hk$ index to metallicity changes is high
at all [Fe/H] for hotter stars and also for cooler stars more metal-poor than 
[Fe/H] = -1.0.  It is about three times more sensitive than the $m_{1}$ index 
(see Fig. 9 of Twarog \& Anthony-Twarog 1995).
Baird (1996, hereafter B96) extended $Caby$ photometry to RR Lyrae stars of known metallicity
and showed that the $hk$ index retains good sensitivity even at the hottest phases
of pulsation. It was demonstrated that isometallicity lines formed
in the $hk/(b-y)$ diagram are single valued with respect to both $b-y$ 
and $hk$. Therefore, the $hk/(b-y)$ diagram gives consistent metallicities
throughout a star's pulsational cycle, including during rising light and
near maximum light, when $\Delta$$S$ results are unreliable, and so
precise knowledge of light curve phase is unnecessary.  An additional advantage of
the photometric approach is that standard crowded-field techniques can be
used to measure stars even in rich cluster centers.

In this paper we present the results of a new $Caby$ photometric survey of 
131 RR Lyrae stars in $\omega$ Cen, from which metal abundances are derived via 
the $hk$ index. In section 2, we describe the observations and the reduction
procedures. The adopted metallicity calibration procedures are outlined in
section 3. In section 4, we present the results of our metallicity 
determination for field RR Lyrae stars and $\omega$ Cen RR Lyrae stars,
with a comparison with the previous $\Delta$$S$ measurements. 
Finally, in section 5 we discuss the impact of our new metallicity measurements 
on the M$_{V}$(RR) - [Fe/H] and period-shift - [Fe/H] relations.
The color-magnitude diagram resulting from the $Caby$ photometry, and a discussion
of the metallicity distribution of giant branch stars will be presented in a 
future paper of this series.

\section{OBSERVATIONS AND DATA REDUCTIONS}

All the observations were made using the CTIO 0.9 m telescope and Tektronix 2048
No. 3 CCD during three nights of an observing run in March 1997. We covered 
$\omega$ Cen in a 3 $\times$ 3 grid and observed one sequence of this
grid per each night. 
The field size of each grid point was 13$\arcmin$.6 $\times$ 13$\arcmin$.6 with
a pixel scale of 0$\arcsec$.40. Our program field, centered on the cluster, 
covers approximately 40$\arcmin$ $\times$ 40$\arcmin$ which roughly
corresponds to the area enclosed within the half tidal radius
of $\omega$ Cen. Typical exposure times were 1400 s for $Ca$, 360 s for $b$,
and 180 s for $y$, with the CCD being read out simultaneously
through all four amplifiers, using an Arcon CCD controller.
The observation log for the program fields is presented in Table 1.
Two to four frames were taken in each band and each field. 

The frames were calibrated from twilight or dawn sky flats and zero-level 
exposures, using the IRAF QUADPROC routines. Calibration frames were made 
by combining several individual exposures. All exposure times were
sufficiently long that the center-to-corner shutter timing error was negligible.
These procedures produced object frames with the sky flat to better than 
1\% in all filters. The IRAF routine COSMICRAYS was used to remove nearly
all of the cosmic ray events in each frame, with conservative parameters
set to avoid corrupting the stellar profiles. 

Photometry of $\omega$ Cen stars was accomplished using DAOPHOT II and  ALLSTAR
(Stetson 1987, 1995). For each frame, a Moffat function PSF, varying cubically
with radial position, was constructed using 100 to 200 
bright, isolated, and unsaturated stars. The PSF was improved iteratively by 
subtracting faint nearby companions of the PSF stars. Aperture corrections were  
calculated using the program DAOGROW (Stetson 1990). The final aperture correction 
were made  
by adjusting the ALLSTAR magnitude of all stars by the weighted mean of the
difference between the total aperture magnitude and the profile-fitting
ALLSTAR magnitude for selected stars (e.g., PSF stars). After the aperture correction, 
we used DAOMATCH/DAOMASTER (Stetson 1992) to match stars of all frames
covering the same field, and derived the average instrumental magnitude and
colors on the same photometric scale. For each frame, the magnitude offset with
respect to each master frame in $Ca$, $b$, and $y$ was calculated, and photometry
for the two to four frames for the same field was transformed to a common instrumental
system. 

On each night, five to seven standards from the list of 
Twarog \& Anthony-Twarog (1995) were observed, and due to the small sample size 
the results for each night were combined.  Comparison of the instrumental 
magnitudes for the final 15 observations in each filter with 
the standard values allowed the construction of linear
transformations for the observed $y$, $b-y$, and $hk$ magnitudes from the
instrumental to the standard system. The standard stars observed cover a color
range of 0.1 - 0.7 and 0.2 - 1.4 for $b-y$ and $hk$, respectively, and an
air mass range of 1.0 - 1.6. Extinction coefficients for all the filters were
determined by a series of standard stars over a wide range of airmass. The final
transformation equations were obtained by a linear least-square fit. They are
        $$b-y = 0.956(b-y)_{i} - 0.013,$$
        $$hk = 0.891hk_{i} -1.013,$$
        $$y = y_{i} + 0.026(b-y)_{i} - 5.007,$$
where $b-y$, $hk$, and $y$ are the color indices and visual magnitude in the 
standard $Caby$ system, $(b-y)_{i}$, $hk_{i}$, and $y_{i}$ refer to 
instrumental magnitudes corrected for extinction.
No other trends in the residuals were noticeable, and therefore no additional
terms in the transformation equations appear to be necessary.
The calibration equations relate observed to standard values for $y$, $b-y$, 
and $hk$ with standard deviations of 0.01, 0.01, and 0.02, respectively.
During the observing runs, six field RR Lyrae ``standard'' stars (four RR$ab$ stars 
and two RR$c$ stars) were observed in order to make a comparison between our result 
and that of B96, as discussed in the next section.
Throughout, we have corrected for reddening using the reddening ratios,
$E(b-y)/E(B-V)$ = 0.75, $E(hk)/E(b-y)$ = -0.1, adopted by B96.

\section{METALLICITY CALIBRATION}

B96 successfully provided the [Fe/H] vs. $hk_{o}$ calibrations for two 
values of $(b-y)_{o}$ = 0.15 and 0.30, from eight RR$ab$ stars and two RR$c$ stars.
Using these relations, it is possible to determine the metallicity of any
RR Lyrae star for which there is $Caby$ photometry at either of these colors.
However, in order to find the metallicity of RR Lyrae stars at arbitrary phase,
it is necessary to find the relations between [Fe/H] and $hk_{o}$ for various 
values of $(b-y)_{o}$ and ultimately produce a set of isometallicity lines that 
are continuous across the full range of $(b-y)_{o}$. In addition to two 
calibrations for $(b-y)_{o}$ = 0.15 and 0.30, Baird \& Anthony-Twarog (1999) 
added a new set of calibrations for a more complete grid of $(b-y)_{o}$ values
[i.e., $(b-y)_{o}$ = 0.20, 0.25, and 0.35] from high-quality photometric
data for 14 RR$ab$ stars, combined with previous data from B96.
As did B96, the metallicity values of Layden (1994) were adopted 
because they provide a uniform set of values for all the field RR$ab$
stars, and they are based on the Zinn \& West (1984; hereafter ZW) metallicity
scale for Galactic globular clusters. Layden's (1994) metallicities for
RR$ab$ stars are based on the relative strengths of the Ca II K line and the
H$_{\delta}$, H$_{\gamma}$, and H$_{\beta}$ Balmer lines. The [Fe/H] values 
of the RR$c$ stars were adopted from Kemper (1982) and transformed to the ZW scale 
with Layden's (1994) equation. In the following discussion, we will denote 
[Fe/H]$_{spec}$ as the metallicity measured spectroscopically for RR$ab$ 
and RR$c$ standard stars used in our calibration.
The final [Fe/H]$_{hk}$ vs. $hk_{o}$ relations were obtained by a straight
line fit (Baird \& Anthony-Twarog 1999). They are
   $$[Fe/H]_{hk} = 8.11hk_{o} - 3.37 \;(\sigma_{rms} = 0.110) \quad for \;\;(b-y)_{o} = 0.15,$$
   $$[Fe/H]_{hk} = 7.75hk_{o} - 3.28 \;(\sigma_{rms} = 0.055) \quad for \;\;(b-y)_{o} = 0.20,$$
   $$[Fe/H]_{hk} = 7.45hk_{o} - 3.36 \;(\sigma_{rms} = 0.035) \quad for \;\;(b-y)_{o} = 0.25,$$
   $$[Fe/H]_{hk} = 6.44hk_{o} - 3.36 \;(\sigma_{rms} = 0.040) \quad for \;\;(b-y)_{o} = 0.30,$$
   $$[Fe/H]_{hk} = 5.06hk_{o} - 3.13 \;(\sigma_{rms} = 0.074) \quad for \;\;(b-y)_{o} = 0.35.$$
The $\sigma_{rms}$ are root mean square deviations calculated in the sense 
[Fe/H]$_{spec}$ - [Fe/H]$_{hk}$, where [Fe/H]$_{hk}$ is the value calculated from 
the observed $hk_{o}$ values using the above relations. The $\sigma_{rms}$ values
are highest at the extreme colors, i.e., at $(b-y)_{o}$ = 0.15 and 0.35, where
the number of calibrating points is lowest.

Figure 1 shows the derived [Fe/H]$_{hk}$ vs. $hk_{o}$ relations for five values
of $(b-y)_{o}$, along with the photometric indices for 14 field RR Lyrae stars 
that define the relations. At warmer temperatures, the sensitivity of $hk_{o}$ to
[Fe/H]$_{hk}$ drops, and the slope in a [Fe/H]$_{hk}$ vs. $hk_{o}$ relation
becomes steeper. When stars get hotter than $(b-y)_{o}$ = 0.25, the slopes of 
the [Fe/H]$_{hk}$ vs. $hk_{o}$ relations are nearly the same, indicating that
[Fe/H]$_{hk}$ is a function of $hk_{o}$ only, as suggested by B96. 
$Caby$ photometry is useful for stars as blue as $(b-y)_{o}$ = 0.10, but at
higher temperatures the sensitivity of $Caby$ photometry to metallicity will 
certainly decrease, and the contamination by the H$_{\epsilon}$  line
line should become quite substantial (Baird \& Anthony-Twarog 1999).

We have calculated the metal abundance of the RR Lyrae stars in the field and 
$\omega$ Cen using the above [Fe/H]$_{hk}$ vs. $hk_{o}$ relations. Additional
[Fe/H]$_{hk}$ vs. $hk_{o}$ relations were derived as necessary by interpolating 
these relations within the range of 0.15 $<$ $(b-y)_{o}$ $<$ 0.35. However, for 
stars with $(b-y)_{o}$ $<$ 0.15, which lie outside the limits of the current 
[Fe/H]$_{hk}$ vs. $hk_{o}$ relations, we applied the relation for $(b-y)_{o}$ = 0.15
because at these warm temperatures the isometallicity lines are horizontal in the 
$hk_{o}/(b-y)_{o}$ diagram, as described above. We estimate that, with these
procedures, introduced uncertainties will be less than 0.1 dex at any point of 
the $hk_{o}/(b-y)_{o}$ diagram.

\section{RESULTS}

\subsection{Field RR Lyrae Stars}

To check the validity of our [Fe/H]$_{hk}$ calibration, we will compare our observations
of field RR Lyraes with those by B96. Our measured values of [Fe/H]$_{hk}$
for field RR Lyrae stars are listed in Table 2, and comparison between spectroscopic
metallicity, [Fe/H]$_{spec}$ (B96), and our [Fe/H]$_{hk}$ are shown in
Figure 2. The [Fe/H]$_{spec}$ and the reddening of the stars was taken
from Table 1 of B96. For the RR$ab$ stars, our [Fe/H]$_{hk}$ 
is in excellent agreement with [Fe/H]$_{spec}$, with an rms scatter of 0.12 dex.  
On the other hand, the [Fe/H]$_{hk}$ for the two RR$c$ stars show larger scatter
than that for the RR$ab$ stars. For V535 Mon, $(b-y)_{o}$ is very small, and the 
sensitivity of $hk$ index is less than at redder colors, which might account for
at least some of the discrepancy.  The reason for the large deviation of AU Vir
is not clear, however it anticipates the difficulties we have with the 
$\Delta$$S$ measurements for the $\omega$ Cen RR$c$ stars, discussed below in 
sections 4.3 and 4.5.

Using the CTIO 4 m Telescope in 1997 December, Walker (1999) observed 14 $Caby$ 
standard stars and three RR$ab$ stars (U Lep, RY Col, HH Pup) from B96's
list. We reduced this data in the same way as described above,  and list the
derived [Fe/H]$_{hk}$ for the RR Lyrae stars  in Table 3.  These stars are also
plotted in Fig. 2, and are in excellent agreement both with our 0.9 m results and
with Fe/H]$_{spec}$. For all our RR$ab$ stars and those of Walker (1999), the rms 
scatter of [Fe/H]$_{hk}$ corresponds to 0.10 dex.

\subsection{$\omega$ Cen RR Lyrae Stars}

In our program field of $\omega$ Cen, we measured 131 RR Lyrae stars, consisting
of 74 RR$ab$ and 57 RR$c$ stars, which can be compared to the total of 180 $\omega$ 
Cen RR Lyrae stars known to date (Hogg 1973; Kaluzny et al. 1997b). 
For each RR Lyrae star we obtained two to four points of $b-y$ and $hk$, and 
dereddened using the reddening law stated in section 2. We adopted the reddening 
value E($B-V$) = 0.12 from Harris (1996). BDE used E($B-V$) = 0.11 which is
essentially identical to the independent work of Dickens \& Saunders (1965).
Whitney et al. (1998) adopted E($B-V$) = 0.15 for their analysis of the hot
stellar population of the $\omega$ Cen. However, the effect of a small uncertainty
in E($B-V$) values is negligible for our metallicity determination 
($\Delta$[Fe/H] $<$ 0.02 dex). Additional correction for the interstellar
contribution to the K line was ignored ($\Delta$[Fe/H] $\sim$ 0.03 dex, see GTO).
These both affect only the mean cluster [Fe/H] value, not the star-to-star scatter.
Table 4 lists the dereddened values, $(b-y)_{o}$ and $hk_{o}$, and their
photometric errors for each RR Lyrae star.

After obtaining the individual values of [Fe/H]$_{hk}$ for each 
RR Lyrae star, we calculated the mean value of the [Fe/H]$_{hk}$ by weighting with 
the photometric error of the $hk$ value.  A number of data points tagged as poor 
measurements were rejected, and some data points that showed a large deviation 
from their isometallicity line in the $hk_{o}/(b-y)_{o}$ diagram were also excluded. 
Table 5 lists our final weighted mean [Fe/H]$_{hk}$ values in column (3). Column 
(5) is the number of independent measurements 
used in the calculation of the mean [Fe/H]$_{hk}$. For the error of the
mean [Fe/H]$_{hk}$ value, we adopted the standard deviation of the mean of
the individual [Fe/H]$_{hk}$ measures. This error is listed as $\sigma$$_{[Fe/H]}$
in the column (4). For stars with only one data point, their $\sigma$$_{[Fe/H]}$ values
have been set to blank. The typical value of $\sigma$$_{[Fe/H]}$ corresponds to about
0.20 dex.
For those stars where the scatter is larger than typical, it is not 
clear whether this is due to observational error, or to some small non-repeatability 
and/or phase dependence in the $hk_{o}/(b-y)_{o}$ diagram as suggested by B96. 
We do not have sufficient observations per star to clarify this, and encourage more
observations of the field ``standard stars''.\footnote{
During the rapid rise to maximum of RR$ab$ stars, one may question whether the effect of 
the sequence of exposure times for $Ca$, $b$, and $y$ over more than half an hour will
cause errors in the metallicity determinations. However, for a few identified data
points on the rising branch, we did not find severe deviations from other data points
in the $hk_{o}/(b-y)_{o}$ diagram. Furthermore, since our [Fe/H]$_{hk}$ corresponds to
the mean of the individual measures and the typical error of the mean [Fe/H]$_{hk}$
value is small (about 0.2 dex), this effect should be negligible.}
As a reference, we estimate the typical values
of frame-to-frame scatter of HB stars as 0.02 and 0.03 mag for $b-y$ and $hk$, respectively.
This scatter of $hk$ corresponds to an error of less than 0.20 dex in [Fe/H], at any $b-y$.

\subsection{Comparison with Previous $\Delta S$ Observations}

Among the 131 RR Lyrae stars in our $\omega$ Cen field, 56 stars are in common with the
previous $\Delta S$ observations of BDE and GTO, and we make a comparison
between these values, [Fe/H]$_{\Delta S}$
[column (6) of Table 5], and those from our $Caby$ photometry, [Fe/H]$_{hk}$
[column (3) of Table 5]. Most values of [Fe/H]$_{\Delta S}$ 
come from BDE, but for a few stars also observed by GTO, new values have been 
calculated by averaging the measurements of BDE and GTO. All [Fe/H]$_{\Delta S}$
values have been corrected to the ZW metallicity scale using the relation obtained 
by Layden (1994) (i.e., [Fe/H]$_{ZW}$ = 0.90[Fe/H]$_{\Delta S}$ - 0.34) in order
that all the [Fe/H] data is placed on a consistent metallicity scale. Figure 3 
illustrates the residuals in the sense [Fe/H]$_{hk}$ - [Fe/H]$_{\Delta S}$ 
as a function of [Fe/H]$_{\Delta S}$. The closed circles are RR$ab$ stars
while open circles are RR$c$ stars. The larger symbols represent stars with
smaller observational error ($\sigma$$_{[Fe/H]}$ $\leq$ 0.2 dex) in [Fe/H]$_{hk}$. 
It is apparent that a significant difference between [Fe/H]$_{hk}$ and 
[Fe/H]$_{\Delta S}$ is present in a manner which is metallicity dependent.\footnote{
Since Freeman \& Rodgers (1975) used a bigger telescope,
at higher dispersion, than did BDE, it is worth to make a comparison between
our result and that of Freeman \& Rodgers. However, we found the rms scatter between
these two metallicities for 16 RR$ab$ stars to be still large (0.37 dex).}
The residuals for the RR$c$ and RR$ab$ stars appear similar, although
the [Fe/H]$_{hk}$ for most RR$c$ stars is metal-rich compared to [Fe/H]$_{\Delta S}$.

In order to more clearly see metallicity differences between [Fe/H]$_{hk}$ and 
[Fe/H]$_{\Delta S}$ in the $hk_{o}/(b-y)_{o}$ diagram, we introduce $hk_{o, \Delta S}$,
which is the expected value of $hk_{o}$ from [Fe/H]$_{\Delta S}$, and so construct a
$hk_{o, \Delta S}/(b-y)_{o}$ diagram.  We calculate $hk_{o, \Delta S}$ by inserting 
[Fe/H]$_{\Delta S}$ into
inverse equations of our final [Fe/H] vs. $hk_{o}$ relations. For the calculation
of this $hk_{o, \Delta S}$, we retained our observed value of $(b-y)_{o}$.
In Figure 4, we compare our observed $hk_{o}/(b-y)_{o}$ diagram with 
$hk_{o, \Delta S}/(b-y)_{o}$ for 56 RR Lyrae stars. In each diagram, we present schematic
isometallicity lines, which were made from five [Fe/H] vs. $hk_{o}$ relations with
step size of 0.5 dex. It should be noted that our observed
$hk_{o}$ distribution for the RR$ab$ stars is slightly more compressed than that of 
$hk_{o, \Delta S}$ for all $(b-y)_{o}$. In the case of the RR$c$ stars and for
some RR$ab$ stars with $(b-y)_{o}$ $<$ 0.2, the distribution of $hk_{o}$ is shifted
in the metal-rich direction, by about 0.5 dex in the mean, compared to that of 
$hk_{o, \Delta S}$.
These comparisons confirm that there are systematic differences between [Fe/H]$_{hk}$ 
and [Fe/H]$_{\Delta S}$.

\subsection{Comparison with Other Metallicity Determinations}

For 48 RR$ab$ stars in $\omega$ Cen, Jurcsik (1998, hereafter J98) determined the
empirical [Fe/H] values from the light-curve parameters using the observations
of Kaluzny et al. (1997b). 
Comparing the empirical [Fe/H] values with the $\Delta S$ measurements of 
BDE and GTO samples for RR$ab$ stars, J98 found significant 
discrepancies and suspected that the $\Delta$$S$ data of BDE and GTO
were inaccurate. Schwarzenberg-Czerny \& Kaluzny (1998; hereafter SK98) 
independently compared their empirical [Fe/H] with the $\Delta S$ metallicites, 
[Fe/H]$_{\Delta S}$, of BDE for 11 RR$ab$ stars and revealed no obvious 
correlation between their empirical [Fe/H] and [Fe/H]$_{\Delta S}$. These results
encouraged us to check for consistency between our [Fe/H]$_{hk}$ and the 
empirical [Fe/H] values of J98 and SK98.

Using the list of 47 RR$ab$ stars employed by J98, we found
the rms scatter between the metallicities of J98, [Fe/H]$_{J}$,
and [Fe/H]$_{\Delta S}$ for 23 RR$ab$ stars turned out to be large (0.48 dex),
whereas that between [Fe/H]$_{J}$ and our [Fe/H]$_{hk}$ for 47 RR$ab$ stars
is much smaller (0.23 dex). Comparing the empirical metallicities independently
obtained by SK98, [Fe/H]$_{SK98}$, with the $\Delta S$ observations for the 11
RR$ab$ stars in common, significant discrepancies were found with a 0.44 dex
rms scatter. However, from the comparison between [Fe/H]$_{SK98}$ and our
[Fe/H]$_{hk}$ for 10 RR$ab$ stars, the scatter reduced to 0.28 dex rms.
In summary, both the empirical metallicities obtained from J98
and SK98 show larger deviation from [Fe/H]$_{\Delta S}$ of BDE and GTO than 
they do from from our photometric [Fe/H]$_{hk}$. Considering the assumed accuracy
of the empirical metallicities as 0.10 - 0.15 dex (Jurcsik \& Kov\'acs 1996; 
J98 and references therein), this is strong evidence that the 
$\Delta S$ measurements of BDE and GTO are subject to larger errors than the
authors state.

\subsection{Metallicity Differences between [Fe/H]$_{hk}$ and [Fe/H]$_{\Delta S}$}

What causes the systematic discrepancies between our [Fe/H]$_{hk}$ and the 
[Fe/H]$_{\Delta S}$ of BDE and GTO?   
We checked that there was no color dependency, which might be the case if
our transformations as a function of color were incorrect.  We also
found that metallicity residuals 
between [Fe/H]$_{hk}$ and [Fe/H]$_{\Delta S}$ showed a similar pattern
for the inner and outer regions of our program field, demonstrating that there
is no dependency on image crowding.

We discuss supporting evidence for there being large errors in the 
$\Delta S$ measurements of BDE and GTO. First of all, the excellent
agreement between [Fe/H]$_{hk}$ and [Fe/H]$_{spec}$, which is compatible to 
the $\Delta S$ measurements (see Layden 1994), of four field RR$ab$ stars 
provides the most positive evidence of the accuracy of the present work and the large
error of the $\Delta S$ results of BDE and GTO (see section 4.1).
Second, there are non-negligible discrepancies between the 
results of BDE and GTO.   GTO claimed that the internal errors are about 0.2 dex
in both BDE and GTO $\Delta S$ measurements, and their system is thus not far 
from the standard $\Delta S$ system. However, as GTO already noted, a few stars
(V32, V39, and V72) show large deviation (more than 0.5 dex) with BDE's results,
probably, due to observation at phases far from the minimum (see Fig. 2 of GTO).
Furthermore, the rms scatter of the mean difference of the $\Delta S$ measurements
between BDE and GTO corresponds to 0.34 dex, certainly not negligible.
Third, J98 obtained an unexpectedly large (0.52 dex) rms scatter
between her empirical metallicity values and $\Delta S$ metallicity of BDE,
but comparing empirical data with GTO's observations, a smaller 0.38 dex rms 
scatter was obtained. Therefore, it is suspected that the $\Delta S$ measurements
for $\omega$ Cen, especially BDE's data, are inaccurate.
Both of the empirical metallicities obtained from J98 and SK98 show
smaller rms scatter in our [Fe/H]$_{hk}$ than in $\Delta S$ metallicity,
[Fe/H]$_{\Delta S}$ (see section 4.4). This suggests that our [Fe/H]$_{hk}$
are more accurate than the [Fe/H]$_{\Delta S}$ of BDE and GTO.
Fourth, despite a more extensive sample than that of previous $\Delta S$ 
observations, the relation between magnitude and metallicity of our observations
show a smaller scatter.  We also note consistency with the model predictions of 
Lee (1991) (see section 5.1 and Fig. 6). Finally, as we will see in section 4.6,
the metallicity distribution of our observations for RR$ab$ stars is more
consistent with that of the giant stars of Suntzeff \& Kraft (1996, hereafter SK),
rather than that of the previous $\Delta S$ observations. It was also suggested by 
SK that the large population of very metal-poor stars found from $\Delta$$S$ 
measurements is incorrect.  According to stellar evolution theory,
the RR Lyrae stars are an intrinsically abundance-biased population due to the
low probability that the extremely metal-rich (-poor) red (blue) HB stars evolve
through the instability strip (Lee \& Demarque 1990). Therefore, it is
unreasonable to expect that the metallicity distribution of RR Lyrae stars is
wider than that of their progenitor stars.  Consequently, we conclude
that the systematic discrepancies between our [Fe/H]$_{hk}$ and the [Fe/H]$_{\Delta S}$
of BDE and GTO are caused by the large uncertainties of the $\Delta S$ measurements.

Finally, we discuss the difference in metallicity distribution in our results between
the RR$ab$ and RR$c$ stars. 
Our distribution of metal-rich RR$c$ stars is
difficult to understand from the standpoint of standard metal-rich HB evolutionary
tracks, which do not penetrate into the hotter regions of the instability strip
(Lee \& Demarque 1990). While no definite resolution on the disagreement in the
metallicity can be offered, we suggest that the metal enhancement of RR$c$
stars may be due to the possible contamination of Ca II H by H$_{\epsilon}$. 
For hotter stars, inclusion of the H$_{\epsilon}$ feature will weaken the metallicity
effect because the weakening of the Ca II H line can be partially compensated by
the growth of the Balmer line (Anthony-Twarog et al. 1991). Although we used the 
[Fe/H] vs. $hk_{o}$ relation at $(b-y)_{o}$ = 0.15 for stars with $(b-y)_{o}$ $<$ 0.15
(see section 3), we should treat this data with caution until the
[Fe/H] vs. $hk_{o}$ relations at higher temperatures are confirmed. Furthermore, because 
the [Fe/H] vs. $hk_{o}$ relation at high temperature [e.g., $(b-y)_{o}$ = 0.15],
as shown in Fig. 1, does not extend to metallicity higher than [Fe/H] = -1.0,
the metal-rich end of the RR$c$ stars may also be suspect. More calibration data
for RR$c$ stars will be needed to resolve this problem. 
For this reason, we regard our results for RR$c$ stars to be tentative, and we will
restrict our analysis to RR$ab$ stars in the following discussions.

\subsection{The Metallicity Distribution}

Considering the homogeneity and large sample size of the present database,
it is worthwhile to investigate the metallicity distribution of RR Lyrae
stars and compare it with the earlier results for RR Lyrae stars and
giant stars (BDE; Dickens 1989; Norris et al. 1996; SK).
However, care must be taken when comparing the metallicity distribution
of RR Lyrae stars and giant stars, since as mentioned above RR Lyrae
stars are intrinsically an abundance-biased sample due to the failure of the
extremely metal-rich red HB stars to penetrate into the
instability strip. Furthermore, the frequency of RR Lyrae stars at a given
metallicity depends on the HB morphology as well as the metallicity distribution
of the underlying stellar population (e.g., Lee 1992 and Walker \& Terndrup
1991 for RR Lyrae stars in Baade's window).
Figure 5 presents the metallicity distributions for the RR$ab$ stars and
giant stars. All of the earlier studies and our results agree on the
non-Gaussian shape of the metallicity distribution which contains a sharp rise 
from the low metallicity side, a modal value of [Fe/H] $\approx$ -1.8 and a 
tail of metal-rich stars reaching at least [Fe/H] $\approx$ -0.9. For the
161 - star bright giant (BG) sample and the 199 - star subgiant 
(SGB) sample of SK (Fig. 5c),
the metallicity distribution is narrower than that of
34 - star RR$ab$ sample obtained from the $\Delta$$S$ method (Fig. 5a). 
SK suggested that the large population of very metal-poor stars found in
the $\Delta$$S$ measurement is due to the large rms error of 0.4 dex of
the old $\Delta$$S$ study and probably the strong low-metallicity tail of the
error distribution is spurious. The more complete sample of RR$ab$ stars from our
$hk$ method (Fig. 5b) also shows the paucity of very metal-poor stars.
Consequently, contrary to the case of the $\Delta S$ observations, the range 
of the metallicity distribution of our observations for RR$ab$ stars is 
consistent with that of the giant stars of SK.

While a detailed discussion of the origin of the abundance distribution is outside
the scope of this paper, we wish to point out that in the analysis of a $B,V$ CMD
for $\omega$ Cen containing 130,000 stars, Lee et al. (1999) found several distinct
red giant branches (RGBs). They also showed from population models that 
the most metal-rich RGB is about 2 Gyr younger than the dominant metal-poor component, 
suggesting that $\omega$ Cen has enriched itself over this timescale.  
An extensive study of the metallicity distribution for an homogeneous and nearly
complete sample of $\omega$ Cen giant branch stars, now underway, will
place this result on a firmer footing.  The RR Lyrae stars, being clearly representatives
of the oldest populations in $\omega$ Cen, will be an important part of any enrichment
model for the cluster.

\section{DISCUSSION}
\subsection{The M$_{V}$(RR) - [Fe/H] Relation}

Given our homogeneous metallicity measurements for nearly the whole sample of
$\omega$ Cen RR Lyrae stars, we can turn to a discussion of the magnitude-metallicity
relation.   We will use the intensity mean magnitude values,
$<V>$, given in BDE and Kaluzny et al. (1997b). For the stars whose $<V>$ values are
available from both sources the mean values have been adopted. 
Column (7) of Table 5 lists the $<V>$ for each RR Lyrae stars. While the photometry
of BDE is restricted to the outer region of the cluster, that of Kaluzny et al. (1997b)
covers a larger area including the central region from their extensive observations
(Kaluzny et al. 1996, 1997a). Although Kaluzny et al. (1997b) have claimed
field-to-field differences on the level of a few hundredth's of magnitude and
uncertainties when combining photometry obtained in different fields for the same
variables, we adopted the averaged value of magnitude for stars with multiple
entries in their Table 1. From intercomparison of 66 RR Lyrae stars between BDE
and Kaluzny et al. (1997b), we found a zeropoint offset of $<V>$ as -0.03 $\pm$ 0.06
in the sense BDE minus Kaluzny et al. (1997b). However, considering the intrinsic
spread (or scatter) and random error of $<V>$ for $\omega$ Cen RR Lyrae stars,
this small offset would have only a small or negligible effect on the discussion of the
M$_{V}$(RR) - [Fe/H] relation.  We are presently reducing $BV$ photometry for
$\omega$ Cen RR Lyrae stars, which in the future will provide a more homogeneous and
consistent dataset of $<V>$.

Using the data in Table 5, the observed correlation
between M$_{V}$(RR) and [Fe/H] is presented in Figure 6, where panel (a) is based on
[Fe/H] determined by the previous $\Delta$$S$ measurements while panel (b) is based 
on our new $Caby$ photometry. In the transformation to the absolute magnitude, we 
adopted a distance modulus of V - M$_{V}$ = 14.1 based on the recent evolutionary 
models of M$_{V}$(RR) by Demarque et al. (1999). In Fig. 6b, closed circles are
stars which overlap with the sample of BDE and GTO (i.e., Fig. 6a), while triangles represent
stars only observed in our study. The large symbols are for stars with smaller
observational error ($\sigma$$_{[Fe/H]}$ $\leq$ 0.2 dex) of the [Fe/H]$_{hk}$ with
the same criterion of Fig. 3. It appears that the random errors of [Fe/H]$_{hk}$
are smaller than those of [Fe/H]$_{\Delta S}$ in the M$_{V}$(RR) - [Fe/H] distribution.
In particular, V5 ([Fe/H]$_{\Delta S}$ = -2.32) and V56 ([Fe/H]$_{\Delta S}$ = -1.82),
which are fainter (about 0.2 mag.) than similarly metal-poor RR Lyrae stars in the
M$_{V}$(RR) - [Fe/H]$_{\Delta S}$ diagram, are moved to relatively metal-rich 
[Fe/H]$_{hk}$. Their new metallicities ([Fe/H]$_{hk}$ = -1.35 and -1.26) are more
consistent with their intrinsic luminosity, following a general trend shown in Fig. 6b.

We have superimposed the model correlations of Lee (1991), which were constructed
based on his HB population models under two assumptions regarding the variation of
HB type with metallicity.
The solid (age = 13.5 Gyr) and short-dashed (age = 15.0 Gyr)
lines are for the case that the HB type follows the nonmonotonic behavior with
decreasing [Fe/H] similar to that observed in the Galactic globular cluster system
[see Fig. 3 of Lee (1991)]. Lee (1991) suggested that this nonmonotonic behavior
of HB morphology is perhaps due either to the highly nonlinear relationship between
mass loss and [Fe/H] or to some combination of the effects of mass loss and enhanced
$\alpha$-elements, although the complete understanding is still lacking. The long-dashed 
line is a simple model locus, with fixed mass loss, age, and $\alpha$-elements,
which fails to reproduce the observed nonmonotonic behavior of HB type with 
decreasing [Fe/H]. The sudden upturn in M$_{V}$(RR) of model loci can be
explained by a series of HB population models (see Fig. 5 of Lee 1993), where one
can see how sensitively the population of the instability strip changes with
decreasing [Fe/H]. As [Fe/H] decreases, there is a certain point where the zero
age portion of the HB just crosses the blue edge of the instability strip.
Then, only highly evolved stars from the blue HB can penetrate back into the
instability strip, and the mean RR Lyrae luminosity increases abruptly (Lee 1993).
As shown in Fig. 6b, the correlation predicted from the model loci, including
the sudden upturn in M$_{V}$(RR), agree better with our new
M$_{V}$(RR) - [Fe/H]$_{hk}$ distribution.
Note that the choice of HB evolutionary tracks has little effect on this
conclusion, as Demarque et al. (1999) recently showed that new synthetic
HB models based on evolutionary tracks with improved input physics (Yi et al. 1997) 
produce qualitatively the same results.

Lee (1991) noted that the solid line in the M$_{V}$(RR) - [Fe/H]$_{\Delta S}$
diagram of Fig. 6a does not pass through a few stars at [Fe/H]$_{\Delta S}$
$\approx$ -1.4, and he suspected this deviation is perhaps due either to the 
observational errors or to the zero point uncertainty of the metallicity scale.
Alternatively, he suggested that a better match is obtained by the older model locus
of 15.0 Gyr (i.e., short-dashed line of Fig. 6). In our new diagram of Fig. 6b, we
can see that these deviant stars are now moved to [Fe/H]$_{hk}$ $\approx$ -1.5
and are, therefore, more well matched to the model locus. With our new data,
the best match with the models is expected somewhere between the solid and
dashed lines (i.e., $\sim$ 14.3 Gyr). Note that the absolute ages in these models
are based on the assumption that the mean age of the inner halo clusters is
$\sim$ 14.5 Gyr, thus this result suggests $\omega$ Cen is comparable in age
with other inner halo clusters.

If we remove stars in the range -1.9 $<$ [Fe/H]$_{hk}$ $<$ -1.5, where most of the
variables are believed to be extremely evolved stars (see section 5.4 below), 
then we obtain $\Delta$M$_{V}$(RR)/$\Delta$[Fe/H] = 0.24 $\pm$ 0.04, which is
consistent, to within the errors, with the slopes obtained by LDZ and Lee (1990) 
from the evolutionary models, excluding the clusters in this metallicity range.
Consequently, RR Lyrae stars in $\omega$ Cen and their nonlinear M$_{V}$(RR) -
[Fe/H] relations from our observations provide a strong support for the LDZ and 
Lee (1990) evolutionary models.   This non-linearity, which also implies that the 
relation between period-shift and metallicity is not linear (see section 5.3 below), 
would clarify some of the disagreements with other investigators because
fits of straight lines to different data sets produce significantly different
slopes.

\subsection{The m$_{bol}$ - logT$_{eff}$ Diagram of RR Lyrae Stars}

In order to test more clearly the metallicity dependence of the luminosity of 
RR Lyrae stars, we constructed the bolometric magnitude (m$_{bol}$) -
temperature (T$_{eff}$) diagram for RR$ab$ stars of different metallicities
in Figure 7. Panel (a) and (b) contains 27 and 34 RR$ab$ stars for which
[Fe/H] has been determined from the $\Delta$$S$ method and our $Caby$ photometry,
respectively. For the $B-V$, we used
the equilibrium color defined by Bingham et al. (1984), $(B-V)_{eq}$ = 
${2\over3}<B-V>$ + ${1\over3}(<B>-<V>)$. In the calculations of $T_{eff}$
and bolometric correction, we adopted a color-temperature relation that
has been used in the construction of the Revised Yale Isochrones 
(Green et al. 1987; see Green 1988) for the consistency with the work
of Lee (1991). The color information, $<B-V>$ and $<B>-<V>$, of the
$\omega$ Cen variables was taken from Sandage (1981).

Fig. 7a shows that the relationship between metallicity and bolometric
magnitude is not clear when the metallicities determined by previous
$\Delta$S observations are used (Dickens 1989). Not all the faintest stars are
the most metal-rich stars, and some metal-rich stars are apparently as bright
as the metal-poor stars. On the other hand, as shown in Fig 7b, the metallicity
dependence of RR Lyrae magnitude becomes more distinct when we use our new
[Fe/H]$_{hk}$ metallicities. Now, most metal-rich RR Lyrae stars lie below
(i.e., having fainter magnitude) the RR Lyrae stars that are relatively more
metal-poor.  The cases of V5 and V56 were discussed above in this context. The 
magnitude gap between the metal-rich and metal-poor RR Lyrae stars near 
m$_{bol}$ $\approx$ 14.6 is due to the abrupt increase, in magnitude at 
approximately [Fe/H] = -1.5 (see Fig. 6b).

\subsection{The Period Shift Effect}

If the relationship between M$_{V}$(RR) and [Fe/H]$_{hk}$ is not linear as 
noticed above, we expect a similar correlation between period-shift and [Fe/H]
for $\omega$ Cen RR Lyrae stars. In order to confirm this, we obtained the
period-shifts of $\omega$ Cen RR$ab$ stars at a fixed T$_{eff}$ from the
deviations in the period of each $\omega$ Cen RR$ab$ star from the M3 fiducial
line in the logP - logT$_{eff}$ plane. Periods have been obtained mainly 
from Kaluzny et al. (1997b), but for some stars we adopted the values of BDE.
Column (8) of Table 5 gives the period for each RR Lyrae star.
As we did in section 5.2, we also used
the $(B-V)_{eq}$, calculated from the photometry of Sandage (1981) and
$(B-V)$ - T$_{eff}$ relations of Green et al. (1987) in the calculations of
$T_{eff}$ for M3 RR Lyrae stars. We transformed the observed periods of M3
RR$c$ stars to fundamental periods by adding 0.125 to their logarithms
(Bingham et al. 1984; LDZ) to obtain the logP - log$T_{eff}$ relationship for 
all the M3 RR Lyraes.

The correlation between period-shift, $\Delta$log$P(T_{eff})$, and [Fe/H] is shown 
in Figure 8. Assuming no large differences between P$^{\prime}$
and P even in the case of M$_{bol}$ range ($\sim$ 0.2 mag) for $\omega$ Cen
RR Lyrae stars, we superimposed the model loci of $\Delta$logP$^{\prime}$(T$_{eff}$) 
- [Fe/H] of Lee (1993; see his Fig. 6b) for the comparison with observations.
The P$^{\prime}$ corresponds to the ``reduced period'', which is corrected for the
differing luminosity within the cluster by normalizing to the mean magnitude
of RR Lyrae stars (see LDZ). As shown in Fig. 8a, for the 27 RR$ab$ stars,
whose [Fe/H] has been determined by the $\Delta$$S$ method, there is no distinct
$\Delta$logP(T$_{eff}$) - [Fe/H]$_{\Delta S}$ correlation. It should be noted
that the period-shift values for metal-rich stars show similar or even larger shifts,
compared to metal-poor stars. This metallicity effect of the period-shift is much
smaller than that found in the period-shift - [Fe/H] relationship of Oosterhoff I
and Oosterhoff II clusters, as well as the field RR Lyrae stars covering a similar 
range of metallicity (LDZ; Lee 1990, 1993). However, when we adopt our new metallicity,
[Fe/H]$_{hk}$, a more clear correlation between $\Delta$logP(T$_{eff}$) and [Fe/H]$_{hk}$
emerges (Fig. 8b) which follows the model locus of Lee (1993), 
despite some scatter among the metal-poor stars. This is expected, $\omega$ Cen
RR Lyrae stars should show more scatter than the models for globular cluster system
because post-ZAHB luminosity evolution causes scatter in period-shifts and also
because only single determinations of period and [Fe/H] exists for individual
$\omega$ Cen RR Lyrae stars, whereas those for the clusters represent averages
over many stars (see Lee 1990). As in the M$_{V}$(RR) - [Fe/H]$_{hk}$ diagram, V5 and
V56 now belong to the metal-rich stars in $\Delta$logP(T$_{eff}$) - [Fe/H]$_{hk}$ 
diagram. The RR Lyrae stars having -1.9 $<$ [Fe/H] $<$ -1.5, which are considered
to be highly evolved stars arisen from the bluest HBs, are shifted in period relative
to M3 variables of the same T$_{eff}$ by approximately the same (or even larger)
amounts as the variables in the more metal-poor RR Lyrae stars (see section 5.4
below). 
Consequently, our new correlation between period-shift and [Fe/H]$_{hk}$ shows roughly 
the same trend as the M$_{V}$(RR) - [Fe/H]$_{hk}$ relation, and is more in agreement
with the model locus.

\subsection{Highly Evolved RR Lyrae Stars}

The effect of post ZAHB evolution plays a key role in our
understanding of the M$_{V}$(RR) - [Fe/H] relation and other related problems,
such as the Sandage period-shift effect (see also Lee 1993 and references therein).
In particular, the evolutionary models of Lee (1990) suggest that the RR Lyrae stars
in very blue HB clusters within the range -2.0$<$[Fe/H]$<$-1.6 are highly evolved stars
from the bluest HBs, and have significantly brighter magnitudes and longer periods 
than those near the ZAHB (see also Figs. 6 and 8).
Highly evolved RR Lyrae stars can be identified from a star's 
position in a period (logP) - blue amplitude ($A_{B}$) diagram by comparing them
with RR Lyrae stars in clusters having similar metallicity but with redder HB morphology 
(Jones et al. 1992; Cacciari et al. 1992; Clement \& Shelton 1999).
Assuming that $A_{B}$ depends on [Fe/H] as well as $T_{eff}$ (LDZ; Caputo 1988),
at a fixed metallicity, relative $A_{B}$ values are reliable indicators of 
relative $T_{eff}$. Therefore, highly evolved RR Lyrae stars in $\omega$ Cen can be
detected from a series of logP - $A_{B}$ diagrams covering the range of metallicities.

Figure 9 shows a logP - $A_{B}$ diagram for $\omega$ Cen RR$ab$ stars at three
metallicity groups with [Fe/H]$_{hk}$ $<$ -1.9, -1.9 $\le$ [Fe/H]$_{hk}$ $<$ -1.5,
and [Fe/H]$_{hk}$ $\ge$ -1.5, respectively. The A$_{B}$ values are from Sandage (1981)
as given in column (9) of Table 5. For comparison, we also plotted RR$ab$ stars in 
M15 ([Fe/H] = -2.17; Lee, Demarque, \& Zinn 1994), M3 ([Fe/H] = -1.66), and M4 
([Fe/H] = -1.28) for each metallicity group, respectively, with data from Sandage (1990b).
The solid line represents the fiducial line of the lower envelope to the M3 distribution of
logP - $A_{B}$ (Sandage 1990a). For the most metal-poor (Fig. 9a) and the most metal-rich
stars (Fig. 9c), the majority of $\omega$ Cen RR Lyrae stars do not, respectively,
show deviations from the M15 and M4 variables. This would indicate that the evolutionary
stages of these $\omega$ Cen variables are not significantly different from those
for variables in M15 and M4, respectively. However, most $\omega$ Cen RR$ab$
stars in the range -1.9 $\le$ [Fe/H] $<$ -1.5 are obviously deviant when
compared to the M3 variables. These stars have much longer periods than M3 variables
of similar $A_{B}$, thus most of them are probably evolved with higher luminosities.
In order to provide a reference for highly evolved stars, in Fig. 9b, we include
two field RR$ab$ stars, SU Dra and SS Leo (open triangles), which have similar
metallicity to M3, but are considered to be in a highly evolved and luminous state 
(Jones et al. 1992). The open square represents a M3 RR$ab$ star (V65), which
is in a more advanced evolutionary state than the majority of M3 RR$ab$ stars
(Kaluzny et al. 1998; Clement \& Shelton 1999). Kaluzny et al. (1998) noted two
other highly evolved M3 RR$ab$ stars, V14 and V104. The similarity of all of these
stars to those in $\omega$ Cen confirms that most $\omega$ Cen RR$ab$ stars
in the range -1.9 $\le$ [Fe/H] $<$ -1.5 are in a highly evolved stage of their 
HB evolution.

Recently, Clement \& Shelton (1999) re-examined the logP - $V$ amplitude (A$_{V}$)
relation of RR Lyrae stars in globular clusters of both Oosterhoff types 
by applying the test of Jurcsik \& Kov\' acs (1996) to identify and remove 
Blazhko variables.  They concluded that the logP - A$_{V}$ relation for ``normal'' 
RR$ab$ stars is not a function
of metal abundance, but rather, related to the Oosterhoff type. Along with the discovery
of three bright M3 RR$ab$ stars in a more advanced evolutionary state, 
they also concluded that the Oosterhoff dichotomy has something to do with evolution 
off the ZAHB. This is consistent with our result presented here, and these 
observations provide a support to the LDZ hypothesis that evolution away from the
ZAHB plays a crucial role in the Oosterhoff period dichotomy (see also Lee \& Carney 1999).

\section{SUMMARY AND CONCLUSIONS}

We present new metallicity measurements of 131 RR Lyrae stars in the $\omega$ Cen
using the $hk$ index of the $Caby$ photometric system. From our study, we draw the 
following conclusions:
\\
\indent (1) We provide the most complete and homogeneous metallicity data
to date, with a typical internal error of 0.20 dex,  based on the [Fe/H] 
vs. $hk_{o}$ calibrations of Baird \& Anthony-Twarog (1999).  
\\
\indent (2) For RR Lyrae stars in common with the $\Delta$$S$ observations of
BDE and GTO, we find that our metallicity values,
[Fe/H]$_{hk}$, are systematically deviant from the $\Delta$$S$ metallicities,
[Fe/H]$_{\Delta S}$, whereas the [Fe/H]$_{hk}$ for well observed field RR$ab$ stars
are consistent with previous spectroscopic measurements. With some supporting
evidence, we find that this discrepancy is due to errors in the BDE and GTO
results.\\
\indent (3) The M$_{V}$(RR) - [Fe/H] and period-shift - [Fe/H] relations from 
our observations show a tight distribution with a nearly step function 
change in luminosity near [Fe/H] = -1.5.  This is consistent with the
model predictions of Lee (1991), which suggest that the luminosity of RR Lyrae
stars depends on evolutionary status as well as metallicity.
\\
\indent (4) From a series of logP - $A_{B}$ diagrams at a range of 
metallicities, we also identify highly evolved RR$ab$ stars in the
range of -1.9 $\le$ [Fe/H]$_{hk}$ $<$ -1.5, as predicted from the synthetic
HB models. Therefore, this gives support to LDZ's hypothesis that
evolution away from the ZAHB plays a role in the Oosterhoff dichotomy.

Some work remains to be done in the future. As noted already, because the 
[Fe/H] vs. $hk_{o}$ relation at high temperature [e.g., $(b-y)_{o}$ = 0.15] shown
in Fig. 1 does not extend to metallicities higher than [Fe/H] = -1.0, the metal-rich
calibration for the RR$c$ stars may be suspect. More calibration data are needed 
to resolve this problem. Furthermore, more RR$c$
stars should be observed to check whether there is any difference between the RR$ab$  
and RR$c$ calibrations. Additionally, in order to test the viability of
the field RR Lyrae stars calibration, it would be valuable to observe samples
of RR Lyrae stars in a number of globular clusters with various and well-determined
metallicities. Then we can determine if the calibrations for field RR Lyrae stars
are consistent with those for the globular cluster RR Lyrae stars.
On the theoretical side, it would be useful to study the relationship between [Fe/H]
and $hk$ using synthetic spectra, and in particular, clarify the problem of 
the contamination of Ca II H by H$\epsilon$ for hotter stars. 
Finally, with its distinct advantages such as ease of observations and analysis, 
the $hk$ method should supersede the old $\Delta S$ method in determining the metallicity
of RR Lyrae stars, despite the need for more accurate calibrations.

\acknowledgments
We would like to thank A. Jurcsik and N. Suntzeff for providing electronic copies
of their datasets, and an anonymous referee for a careful review and useful comments.
S.-C.R. is grateful to Suk-Jin Yoon for his helpful efforts in some model calculations.
Support for this work was provided by the Creative Research Initiatives Program 
of Korean Ministry of Science \& Technology, and in part by the Korea Science \& 
Engineering Foundation through grant 95-0702-01-01-3.

\appendix
\section{NOTES ON INDIVIDUAL RR LYRAE STARS}

$V24$.$-$V24 ([Fe/H]$_{hk}$ = -1.86) has a very small period-shift value
[$\Delta$logP(T$_{eff}$) = -0.09], compared with normal RR$ab$ stars (see Fig. 8b).
Considering its light-curve characteristics, such as period (0.4623 day; Kaluzny et al. 1997b)
and blue amplitude (A$_{B}$ = 0.47; Sandage 1981) (see Fig. 9b), and sinusoidal
light-curve shape (Kaluzny et al. 1997b), V24 is likely to be an RR$c$ star.

$V52$.$-$Kaluzny et al. (1997b) suggested that V52, which is the brightest
RR$ab$ star ($<V>$ = 13.95), is actually a BL Her variable.  However, its 
period (0.66 day) is significantly shorter than the 0.75 day,
short period limit found for Pop. II Cepheids (Wallerstein \& Cox 1984).

$V7$, $V116$, $and$ $V149$.$-$Unlike other metal-rich stars, V116 and V149 show brighter 
magnitudes, comparable to those of relatively metal-poor stars (see Fig. 6b). 
Considering the large deviation in the logP - A$_{B}$ diagram (see Fig. 9), V149
is probably a highly evolved RR Lyrae star. According to its period (0.72 day)
and $V$ amplitude, A$_{V}$ (0.54 mag, Kaluzny et al. 1997b), V116 is likely to be in a
similar evolutionary state.
Although the luminosity is not as high as that of V116 and V149, considering its large   
deviation in the logP - A$_{B}$ diagram and metallicity ([Fe/H]$_{hk}$ = -1.46)
close to the boundary for the evolved stars, it is not
unreasonable to consider V7 as a highly evolved star, also.

\clearpage
\begin{deluxetable}{cclcccc}

\setcounter{table}{0}
\tablewidth{0pt}
\tablecaption{$\omega$ Cen Observation Log}
\tablehead{
\colhead{}     & \colhead{}     & \colhead{}     & \multicolumn{3}{c}{Exposures}  & \colhead{} \\
\colhead{}     & \colhead{}     & \colhead{}     & \multicolumn{3}{c}{(number$\times$seconds)}  &
\colhead{} \\
\cline{4-6} \\
\colhead{}  & \colhead{}   & \colhead{Frame Center} & \colhead{}   & \colhead{}   & \colhead{}  &
\colhead{FWHM} \\
\colhead{Observing Date}   &   \colhead{Field}   &   \colhead{(arcmin)}   &
\colhead{$y$}        &   \colhead{$b$}   &  \colhead{$Ca$}   &
\colhead{(arcsec)}}
\startdata
      27/28 March 1997  &  F1  & 18 SE & 1$\times$180  & 1$\times$360  &   1$\times$1200   &    1.9  \nl
                       &  F2  & 13 E  & 1$\times$180  & 1$\times$360  &   1$\times$1400   &    1.8  \nl
                       &  F3  & 18 NE & 1$\times$180  & 1$\times$360  &   1$\times$1400   &    1.7  \nl
                       &  F4  & 13 N  & 1$\times$180  & 1$\times$360  &   1$\times$1400   &    1.8  \nl
                       &  F5  & cluster center &  1$\times$180  & 1$\times$360  &   1$\times$1400  &  1.6  \nl
                       &  F6  & 13 S  & 1$\times$180  & 1$\times$360  &   1$\times$1400   &    1.5  \nl
                       &  F7  & 18 SW & 1$\times$180  & 1$\times$360  &   1$\times$1400   &    1.6  \nl
                       &  F8  & 13 W  & 1$\times$180  & 1$\times$360  &   1$\times$1400   &    1.6  \nl
                       &  F9  & 18 NW & 1$\times$180  & 1$\times$360  &   1$\times$1400   &    1.5  \nl
\\
      28/29 March 1997  &  F1  & 18 SE & 1$\times$180  & 1$\times$360  &   1$\times$1400   &    2.1  \nl
                       &  F2  & 13 E  & 1$\times$180  & 1$\times$360  &   1$\times$1400   &    2.2  \nl
                       &  F3  & 18 NE & 1$\times$180  & 1$\times$360  &   1$\times$1400   &    2.0  \nl
                       &  F4  & 13 N  & 1$\times$180  & 1$\times$360  &   1$\times$1400   &    1.8  \nl
                       &  F5  & cluster center & 1$\times$180  & 1$\times$360  &   1$\times$1400   &  1.8  \nl
                       &  F6  & 13 S  & 1$\times$180  & 1$\times$360  &   1$\times$1400   &    2.0  \nl
                       &  F7  & 18 SW & 1$\times$180  & 1$\times$360  &   1$\times$1400   &    1.9  \nl
                       &  F8  & 13 W  & 1$\times$180  & 1$\times$360  &   1$\times$1400   &    2.0  \nl
                       &  F9  & 18 NW & 1$\times$180  & 1$\times$360  &   1$\times$1400   &    1.9  \nl
\\
      29/30 March 1997  &  F1  & 18 SE & 1$\times$180  & 1$\times$360  &   1$\times$1400   &    1.7  \nl
                       &  F2  & 13 E  & 2$\times$180  & 2$\times$360  &   2$\times$1400   &    1.6  \nl
                       &  F3  & 18 NE & 1$\times$180  & 1$\times$360  &   1$\times$1400   &    1.6  \nl
                       &  F4  & 13 N  & \nodata  & \nodata  &   \nodata   &    \nodata  \nl
                       &  F5  & cluster center & 1$\times$180  & 1$\times$360  &   1$\times$1400   &  1.6  \nl
                       &  F6  & 13 S  & 1$\times$180  & 1$\times$360  &   1$\times$1400   &    1.6  \nl
                       &  F7  & 18 SW & 1$\times$180  & 1$\times$360  &   1$\times$1400   &    1.6  \nl
                       &  F8  & 13 W  & 1$\times$180  & 1$\times$360  &   1$\times$1400   &    1.5  \nl
                       &  F9  & 18 NW & 1$\times$180  & 1$\times$360  &   1$\times$1400   &    1.7  \nl
\enddata
\end{deluxetable}

\clearpage
\begin{deluxetable}{lcccccc}

\setcounter {table}{1}
\tablewidth{0pt}
\tablecaption{Observations of Field RR Lyrae Stars}
\tablehead{
\colhead{Star}       &      \colhead{$E(B-V)$$^a$}  &
\colhead{$(b-y)_{o}$}        & \colhead{$hk_{o}$}   &
\colhead{$[Fe/H]_{spec}$$^a$}  & \colhead{$[Fe/H]_{hk}$} &   \colhead{Type}}
\startdata
     WY Ant     &  0.014    &  0.334   &  0.274   &  -1.66  &  -1.71  &  ab   \nl
     RY Col     &  0.012    &  0.304   &  0.339   &  -1.11  &  -1.18  &  ab   \nl
     U Lep      &  0.014    &  0.326   &  0.247   &  -1.93  &  -1.83  &  ab   \nl
     HH Pup     &  0.060    &  0.229   &  0.369   &  -0.69  &  -0.52  &  ab   \nl
     V535 Mon   &  0.113    &  0.039   &  0.141   &  -1.64  &  -2.23  &  c    \nl
     AU Vir     &  0.005    &  0.169   &  0.307   &  -2.00  &  -0.90  &  c    \nl
\enddata
\tablenotetext{} {$^a$ From Baird (1996)}
\end{deluxetable}

\clearpage
\begin{deluxetable}{lcccccc}

\setcounter {table}{2}
\tablewidth{0pt}
\tablecaption{Observations of Field RR Lyrae Stars with CTIO 4 m Telescope}
\tablehead{
\colhead{Star}       &      \colhead{$E(B-V)$$^a$}  &
\colhead{$(b-y)_{o}$}        & \colhead{$hk_{o}$}   &
\colhead{$[Fe/H]_{spec}$$^a$}  & \colhead{$[Fe/H]_{hk}$} &   \colhead{Type}}
\startdata
     RY Col     &  0.012    &  0.296   &  0.357   &  -1.11  &  -1.06  &  ab   \nl
     U Lep      &  0.014    &  0.241   &  0.205   &  -1.93  &  -1.83  &  ab   \nl
     HH Pup     &  0.060    &  0.293   &  0.386   &  -0.69  &  -0.77  &  ab   \nl
\enddata
\tablenotetext{} {$^a$ From Baird (1996)}
\end{deluxetable}

\clearpage
\begin{deluxetable}{cccccccccccccccccccc}

\setcounter{table}{3}
\footnotesize
\tablewidth{0pt}
\tablecaption{Photometric Data for $\omega$ Cen RR Lyrae Stars\tablenotemark{*}}
\tablehead{
\colhead{}     & \multicolumn{4}{c}{data1} & \colhead{}  & \multicolumn{4}{c}{data2} & \colhead{}  & \multicolumn{4}{c}{data3}  &
\colhead{} & \multicolumn{4}{c}{data4}  \\
\cline{2-5}      \cline{7-10}   \cline{12-15}    \cline{17-20}  \\
\colhead{Variable}   &   \colhead{$(b-y)_{o}$}   &   \colhead{$\sigma_{b-y}$}   &
\colhead{$hk_{o}$}        &   \colhead{$\sigma_{hk}$}   &   \colhead{}  &
\colhead{$(b-y)_{o}$}   &   \colhead{$\sigma_{b-y}$}   &
\colhead{$hk_{o}$}        &   \colhead{$\sigma_{hk}$}   &    \colhead{}  &
\colhead{$(b-y)_{o}$}   &   \colhead{$\sigma_{b-y}$}   &
\colhead{$hk_{o}$}        &   \colhead{$\sigma_{hk}$}   &    \colhead{}  &
\colhead{$(b-y)_{o}$}   &   \colhead{$\sigma_{b-y}$}   &
\colhead{$hk_{o}$}        &   \colhead{$\sigma_{hk}$}}    
\startdata
   3&    0.257&    0.010&    0.244&    0.009&&    0.255&    0.020&    0.239&    0.018&&    0.189&    0.029&    0.235&    0.026&& \nodata& \nodata& \nodata& \nodata\nl  
   4&    0.303&    0.009&    0.259&    0.009&&    0.301&    0.010&    0.245&    0.010&& \nodata& \nodata& \nodata& \nodata && \nodata& \nodata& \nodata& \nodata\nl  
   5&    0.283&    0.008&    0.312&    0.010&&    0.243&    0.008&    0.260&    0.010&& \nodata& \nodata& \nodata& \nodata && \nodata& \nodata& \nodata& \nodata\nl  
   7&    0.315&    0.010&    0.290&    0.012&&    0.230&    0.010&    0.255&    0.011&& \nodata& \nodata& \nodata& \nodata && \nodata& \nodata& \nodata& \nodata\nl  
   8&    0.291&    0.011&    0.262&    0.014&&    0.323&    0.012&    0.190&    0.013&& \nodata& \nodata& \nodata& \nodata && \nodata& \nodata& \nodata& \nodata\nl  
   9&    0.268&    0.016&    0.262&    0.012&&    0.307&    0.015&    0.292&    0.016&&    0.319&    0.031&    0.282&    0.029&& \nodata& \nodata& \nodata& \nodata\nl  
  10&    0.174&    0.010&    0.196&    0.010&&    0.236&    0.010&    0.229&    0.009&& \nodata& \nodata& \nodata& \nodata && \nodata& \nodata& \nodata& \nodata\nl  
  11&    0.301&    0.011&    0.244&    0.010&&    0.098&    0.007&    0.220&    0.006&&    0.277&    0.010&    0.220&    0.012&& \nodata& \nodata& \nodata& \nodata\nl  
  12&    0.149&    0.008&    0.236&    0.009&&    0.163&    0.009&    0.237&    0.009&&    0.247&    0.009&    0.216&    0.011&& \nodata& \nodata& \nodata& \nodata\nl  
  13&    0.322&    0.025&    0.234&    0.030&& \nodata& \nodata& \nodata& \nodata && \nodata& \nodata& \nodata& \nodata && \nodata& \nodata& \nodata& \nodata\nl
  14&    0.140&    0.007&    0.213&    0.008&&    0.178&    0.009&    0.183&    0.012&& \nodata& \nodata& \nodata& \nodata && \nodata& \nodata& \nodata& \nodata\nl  
  15&    0.319&    0.010&    0.285&    0.011&&    0.223&    0.009&    0.274&    0.010&&    0.322&    0.009&    0.188&    0.010&& \nodata& \nodata& \nodata& \nodata\nl  
  16&    0.154&    0.007&    0.262&    0.007&&    0.156&    0.008&    0.241&    0.012&&    0.150&    0.010&    0.256&    0.011&& \nodata& \nodata& \nodata& \nodata\nl  
  18&    0.308&    0.010&    0.254&    0.010&&    0.273&    0.009&    0.272&    0.010&&    0.272&    0.012&    0.161&    0.012&&    0.331&    0.012&    0.252&    0.012\nl  
  19&    0.215&    0.012&    0.268&    0.011&&    0.128&    0.011&    0.266&    0.011&&    0.196&    0.010&    0.274&    0.013&& \nodata& \nodata& \nodata& \nodata\nl  
  21&    0.084&    0.009&    0.317&    0.008&&    0.161&    0.008&    0.290&    0.009&& \nodata& \nodata& \nodata& \nodata && \nodata& \nodata& \nodata& \nodata\nl  
  22&    0.240&    0.013&    0.230&    0.012&&    0.212&    0.009&    0.241&    0.011&&    0.192&    0.011&    0.179&    0.012&&    0.258&    0.009&    0.238&    0.011\nl
  23&    0.301&    0.013&    0.363&    0.014&&    0.289&    0.015&    0.349&    0.013&&    0.304&    0.013&    0.317&    0.019&& \nodata& \nodata& \nodata& \nodata\nl  
  24&    0.206&    0.013&    0.185&    0.012&&    0.215&    0.010&    0.185&    0.015&&    0.207&    0.012&    0.186&    0.014&& \nodata& \nodata& \nodata& \nodata\nl  
  25&    0.146&    0.012&    0.236&    0.014&&    0.275&    0.017&    0.265&    0.016&&    0.287&    0.019&    0.235&    0.017&& \nodata& \nodata& \nodata& \nodata\nl  
\enddata
\tablenotetext{*}{Table 4 is presented in its complete form in the electronic edition of the
                  Astronomical Journal.\\
                   A portion is shown here for guidance regarding its
                  form and content.}
\end{deluxetable}

\clearpage
\begin{deluxetable}{ccccccccc}

\setcounter {table}{4}
\tablewidth{0pt}
\tablecaption{Metallicities and Photometric Parameters for $\omega$ Cen RR Lyrae Stars}
\tablehead{
\colhead{Variable}       &      \colhead{Type}       &
\colhead{$[Fe/H]_{hk}$}        & \colhead{$\sigma_{[Fe/H]}$}   &   \colhead{N}   &
\colhead{$[Fe/H]_{\Delta S}$$^a$} &   \colhead{$<V>$$^b$}  &   \colhead{Period (days)$^b$}   &
\colhead{$A_{B}$$^c$}}
\startdata
    3&      ab&     -1.54&      0.05&  3&   -1.35&     14.39&   0.8413&  0.92     \nl 
    4&      ab&     -1.74&      0.05&  2&   -1.65&     14.50&   0.6273&  1.29     \nl 
    5&      ab&     -1.35&      0.08&  2&   -2.32&     14.75&   0.5154&  1.28     \nl 
    7&      ab&     -1.46&      0.08&  2&   -1.84&     14.53&   0.7130&  1.13    \nl 
    8&      ab&     -1.91&     0.28&   2&   -1.91&     14.65&   0.5213&  1.39    \nl 
    9&      ab&     -1.49&      0.06&  3&   -1.01&     14.78&   0.5235&  0.97     \nl 
   10&      c&     -1.66&      0.10&   2&   -1.82&     14.51&   0.3750&  0.52     \nl 
   11&      ab&     -1.67&      0.13&  3& \nodata&     14.55&   0.5645&  \nodata  \nl 
   12&      c&     -1.53&      0.14&   3& \nodata&     14.54&   0.3868&  \nodata  \nl 
   13&      ab&     -1.91&   \nodata&  1&   -1.72&     14.48&   0.6691&  1.14     \nl 
   14&      c&     -1.71&      0.13&   2&   -2.22&     14.55&   0.3771&  0.61    \nl 
   15&      ab&     -1.64&      0.39&  3& \nodata&     14.41&   0.8107&  \nodata  \nl 
   16&      c&     -1.29&      0.08&   3& \nodata&     14.56&   0.3302&  0.57    \nl 
   18&      ab&     -1.78&      0.28&  4&   -2.01&     14.52&   0.6217&  1.29    \nl 
   19&      c&     -1.22&      0.05&   3& \nodata&     14.86&   0.2996&  0.54     \nl 
   21&      c&     -0.90&      0.11&   2& \nodata&     14.41&   0.3808&  \nodata  \nl 
   22&      c&     -1.63&      0.17&   4&   -2.35&     14.54&   0.3960&  0.54    \nl 
   23&      ab&     -1.08&      0.14&  3& \nodata&     14.83&   0.5109&  \nodata  \nl 
   24&      ab&     -1.86&      0.03&  3& \nodata&     14.45&   0.4623&  0.47    \nl 
   25&      c&     -1.57&      0.14&   3& \nodata&     14.50&   0.5885&  \nodata  \nl 
   26&      ab&     -1.68&      0.10&  3& \nodata&     14.50&   0.7846&  \nodata  \nl 
   27&      ab&     -1.50&      0.26&  3&   -1.38&     14.75&   0.6157&  0.69     \nl 
   30&      c&     -1.75&      0.17&   3& \nodata&     14.49&   0.4039&  \nodata  \nl 
   32&      ab&     -1.53&      0.16&  2&   -1.55&     14.48&   0.6204&  1.33     \nl 
   33&      ab&     -2.09&      0.23&  3&   -2.02&     14.54&   0.6023&  1.36     \nl 
   34&      ab&     -1.71&   \nodata&  1&   -1.44&     14.49&   0.7340&  0.95     \nl 
   35&      c&     -1.56&      0.08&   4& \nodata&     14.56&   0.3868&  \nodata  \nl 
   36&      c&     -1.49&      0.23&   2&   -2.02&     14.52&   0.3798&  0.55    \nl 
   38&      ab&     -1.75&      0.18&  3&   -1.85&     14.52&   0.7791&  0.75     \nl 
   39&      c&     -1.96&      0.29&   4&   -1.98&     14.57&   0.3934&  0.56    \nl 
   40&      ab&     -1.60&      0.08&  3& \nodata&     14.53&   0.6341&  \nodata  \nl 
   41&      ab&     -1.89&      0.48&  2& \nodata&     14.50&   0.6630&  \nodata  \nl 
   44&      ab&     -1.40&      0.12&  3&   -1.12&     14.70&   0.5675&  1.12     \nl 
   45&      ab&     -1.78&      0.25&  3&   -1.44&     14.53&   0.5891&  1.25    \nl 
   46&      ab&     -1.88&      0.17&  3& \nodata&     14.49&   0.6870&  1.14    \nl 
   47&      c&     -1.58&      0.31&   3& \nodata&     14.29&   0.4851&  0.46     \nl 
   49&      ab&     -1.98&     0.11&   2&   -2.08&     14.63&   0.6046&  1.12    \nl 
   50&      c&     -1.59&      0.19&   3& \nodata&     14.64&   0.3862&  0.53     \nl 
   51&      ab&     -1.64&      0.21&  3& \nodata&     14.53&   0.5742&  \nodata  \nl 
   52&      ab&     -1.42&      0.04&  3& \nodata&     13.95&   0.6604&  \nodata  \nl 
   54&      ab&     -1.66&      0.12&  2& \nodata&     14.42&   0.7729&  0.83    \nl 
   55&      ab&     -1.23&      0.31&  3&   -1.14&     14.77&   0.5817&  1.01    \nl 
   56&      ab&     -1.26&      0.15&  3&   -1.82&     14.75&   0.5680&  1.01    \nl 
   57&      ab&     -1.89&      0.14&  3&   -1.96&     14.48&   0.7946&  0.75    \nl 
   58&      c&     -1.37&      0.18&   3& \nodata&     14.52&   0.3699&  0.25     \nl 
   59&      ab&     -1.00&      0.28&  2& \nodata&     14.76&   0.5185&  \nodata  \nl 
   62&      ab&     -1.62&      0.29&  3& \nodata&     14.48&   0.6199&  \nodata  \nl 
   63&      ab&     -1.73&      0.09&  3&   -2.12&     14.50&   0.8259&  0.57    \nl 
   64&      c&     -1.46&      0.23&   4&   -1.63&     14.56&   0.3446&  0.57    \nl 
   66&      c&     -1.68&     0.34&    4& \nodata&     14.58&   0.4075&  \nodata  \nl 
   67&      ab&     -1.10&   \nodata&  1&   -1.04&     14.69&   0.5644&  1.10    \nl 
   68&      c&     -1.60&      0.01&   2&   -1.95&     14.25&   0.5346&  0.52    \nl 
   69&      ab&     -1.52&      0.14&  2&   -1.91&     14.56&   0.6532&  1.15    \nl 
   70&      c&     -1.94&      0.15&   3& \nodata&     14.53&   0.3907&  0.49     \nl 
   72&      c&     -1.32&      0.22&   3&   -1.88&     14.54&   0.3845&  0.52    \nl 
   73&      ab&     -1.50&      0.09&  3& \nodata&     14.52&   0.5752&  1.31    \nl 
   74&      ab&     -1.83&      0.36&  2&   -1.82&     14.59&   0.5032&  1.49    \nl 
   75&      c&     -1.49&      0.08&   2& \nodata&     14.47&   0.4221&  0.45    \nl 
   76&      c&     -1.45&      0.13&   2&   -2.09&     14.57&   0.3380&  0.42     \nl 
   77&      c&     -1.81&   \nodata&   1&   -1.68&     14.53&   0.4263&  0.48     \nl 
   79&      ab&     -1.39&      0.18&  4&   -1.60&     14.61&   0.6083&  1.30    \nl 
   81&      c&     -1.72&      0.31&   3& \nodata&     14.56&   0.3894&  0.52     \nl 
   82&      c&     -1.56&      0.20&   4& \nodata&     14.51&   0.3358&  0.53     \nl 
   83&      c&     -1.30&      0.22&   2&   -1.76&     14.59&   0.3566&  0.57     \nl 
   84&      c&     -1.47&      0.10&   3&   -0.80&     14.28&   0.5799&  0.81    \nl 
   85&      ab&     -1.87&      0.31&  3&   -1.38&     14.48&   0.7427&  0.86    \nl 
   86&      ab&     -1.81&      0.18&  3& \nodata&     14.54&   0.6479&  \nodata  \nl 
   87&      c&     -1.44&     0.19&    3& \nodata&     14.60&   0.3965&  \nodata  \nl 
   88&      ab&     -1.65&      0.23&  3& \nodata&     14.23&   0.6904&  \nodata  \nl 
   89&      c&     -1.37&      0.28&   3&   -1.38&     14.57&   0.3749&  \nodata  \nl 
   90&      ab&     -2.21&   \nodata&  1& \nodata&     14.53&   0.6034&  \nodata  \nl 
   91&      ab&     -1.44&      0.17&  2& \nodata&   \nodata&  \nodata&  \nodata  \nl 
   94&      c&     -1.00&      0.11&   2& \nodata&     14.76&   0.2539&  0.31    \nl 
   95&      c&     -1.84&      0.55&   3&   -1.57&     14.52&   0.4050&  0.49    \nl 
   96&      ab&     -1.22&   \nodata&  1&   -1.49&   \nodata&   0.6245&  0.89     \nl 
   97&      ab&     -1.56&      0.37&  3& \nodata&     14.53&   0.6919&  \nodata  \nl 
   98&      c&     -1.05&      0.12&   2& \nodata&     14.84&   0.2806&  \nodata  \nl 
   99&      ab&     -1.66&      0.14&  3&   -1.28&     14.30&   0.7661&  1.13     \nl 
  100&      ab&     -1.58&      0.14&  3& \nodata&   \nodata&  \nodata&  \nodata  \nl 
  101&      c&     -1.88&      0.32&   4& \nodata&     14.72&   0.3410&  0.44     \nl 
  102&      ab&     -1.84&      0.13&  4& \nodata&     14.55&   0.6914&  \nodata  \nl 
  103&      c&     -1.92&      0.11&   3& \nodata&   \nodata&  \nodata&  \nodata  \nl 
  104&      ab&     -1.83&      0.18&  4&   -1.79&     14.48&   0.8663&  0.41    \nl 
  105&      c&     -1.24&      0.18&   4& \nodata&     14.70&   0.3353&  0.55    \nl 
  106&      ab&     -1.50&      0.23&  3& \nodata&   \nodata&  \nodata&  \nodata  \nl 
  107&      ab&     -1.36&      0.11&  2& \nodata&     14.82&   0.5141&  \nodata  \nl 
  108&      ab&     -1.93&      0.23&  2& \nodata&     14.62&   0.5945&  \nodata  \nl 
  109&      ab&     -1.51&      0.25&  3& \nodata&     14.45&   0.7439&  \nodata  \nl 
  110&      c&     -2.14&      0.16&   3& \nodata&   \nodata&  \nodata&  \nodata  \nl 
  111&      ab&     -1.66&      0.04&  2& \nodata&     14.46&   0.7630&  \nodata  \nl 
  112&      ab&     -1.81&      0.26&  3& \nodata&     14.60&   0.4743&  \nodata  \nl 
  113&      ab&     -1.65&      0.34&  3& \nodata&     14.60&   0.5733&  \nodata  \nl 
  114&      ab&     -1.32&      0.30&  3&   -2.12&   \nodata&  \nodata&  0.75     \nl 
  115&      ab&     -1.87&      0.01&  2& \nodata&     14.55&   0.6305&  1.18     \nl 
  116&      ab&     -1.27&      0.44&  2& \nodata&     14.27&   0.7200&  \nodata  \nl 
  117&      c&     -1.68&      0.25&   3& \nodata&     14.50&   0.4216&  \nodata  \nl 
  118&      ab&     -1.62&      0.23&  3&   -1.80&     14.43&   0.6116&  \nodata  \nl 
  119&      c&     -1.61&      0.10&   3&   -1.22&     14.66&   0.3059&  \nodata  \nl 
  120&      ab&     -1.39&      0.06&  3& \nodata&     14.74&   0.5486&  \nodata  \nl 
  121&      c&     -1.46&     0.13&    3& \nodata&     14.58&   0.3042&  \nodata  \nl 
  122&      ab&     -2.02&      0.18&  3& \nodata&     14.52&   0.6349&  \nodata  \nl 
  123&      c&     -1.64&      0.01&   2& \nodata&     14.48&   0.4743&  0.49     \nl 
  124&      c&     -1.33&      0.23&   3&   -1.65&     14.60&   0.3319&  0.60    \nl 
  125&      ab&     -1.67&      0.22&  3&   -0.99&     14.59&   0.5929&  1.42    \nl 
  126&      c&     -1.31&      0.13&   3&   -1.96&     14.59&   0.3420&  0.52    \nl 
  127&      c&     -1.59&      0.08&   2& \nodata&     14.62&   0.3053&  0.38     \nl 
  128&      ab&     -1.88&      0.04&  2& \nodata&     14.32&   0.8350&  \nodata  \nl 
  130&      ab&     -1.46&      0.17&  2& \nodata&     14.70&   0.4932&  1.10    \nl 
  131&      c&     -1.56&      0.20&   3& \nodata&     14.50&   0.3921&  \nodata  \nl 
  132&      ab&     -1.91&      0.20&  3& \nodata&   \nodata&  \nodata&  \nodata  \nl 
  134&      ab&     -1.80&      0.41&  3&   -1.83&     14.52&   0.6529&  1.27    \nl 
  135&      ab&     -2.20&   \nodata&  1& \nodata&   \nodata&  \nodata&  \nodata  \nl 
  136&      c&     -1.83&      0.47&   3&   -2.12&   \nodata&  \nodata&  \nodata  \nl 
  137&      c&     -1.19&      0.18&   3& \nodata&     14.53&   0.3342&  \nodata  \nl 
  139&      ab&     -1.46&      0.04&  3&   -2.01&     14.35&   0.6768&  \nodata  \nl 
  141&      ab&     -1.55&      0.36&  2& \nodata&   \nodata&  \nodata&  \nodata  \nl 
  143&      ab&     -1.87&      0.14&  3& \nodata&   \nodata&  \nodata&  \nodata  \nl 
  144&      ab&     -1.71&     0.12&   3&   -1.38&     14.41&   0.8352&  \nodata  \nl
  145&      c&     -1.58&      0.07&   3&   -2.12&     14.56&   0.3732&  \nodata  \nl 
  147&      c&     -1.66&      0.14&   3& \nodata&   \nodata&  \nodata&  \nodata  \nl 
  149&      ab&     -1.21&      0.24&  3& \nodata&     14.42&   0.6827&  1.21    \nl 
  150&      ab&     -1.76&      0.34&  3& \nodata&   \nodata&  \nodata&  \nodata \nl 
  151&      c&     -1.30&      0.24&   3& \nodata&     14.55&   0.4078&  0.42     \nl 
  153&      c&     -1.38&      0.19&   3&   -1.60&     14.55&   0.3863&  \nodata  \nl 
  154&      c&     -1.39&      0.12&   3& \nodata&   \nodata&  \nodata&  \nodata  \nl 
  155&      c&     -1.46&      0.09&   3& \nodata&     14.50&   0.4139&  \nodata  \nl 
  156&      c&     -1.40&      0.04&   2& \nodata&   \nodata&  \nodata&  \nodata  \nl 
  157&      c&     -1.49&      0.10&   3& \nodata&     14.56&   0.4066&  \nodata  \nl 
  158&      c&     -1.25&      0.06&   2& \nodata&     14.59&   0.3673&  \nodata  \nl 
  160&      c&     -1.66&   \nodata&   1&   -2.01&     14.55&   0.3973&  0.52   \nl 
  163&      c&     -1.18&      0.27&   3&   -2.07&     14.56&   0.3132&  0.27    \nl 
\enddata
\tablenotetext{} {$^a$ From Butler et al. (1978) and Gratton et al. (1986).}
\tablenotetext{} {$^b$ From Butler et al. (1978) and Kaluzny et al. (1997b).}
\tablenotetext{} {$^c$ From Sandage (1981).}
\end{deluxetable}

\clearpage

\begin{figure}
\plotone{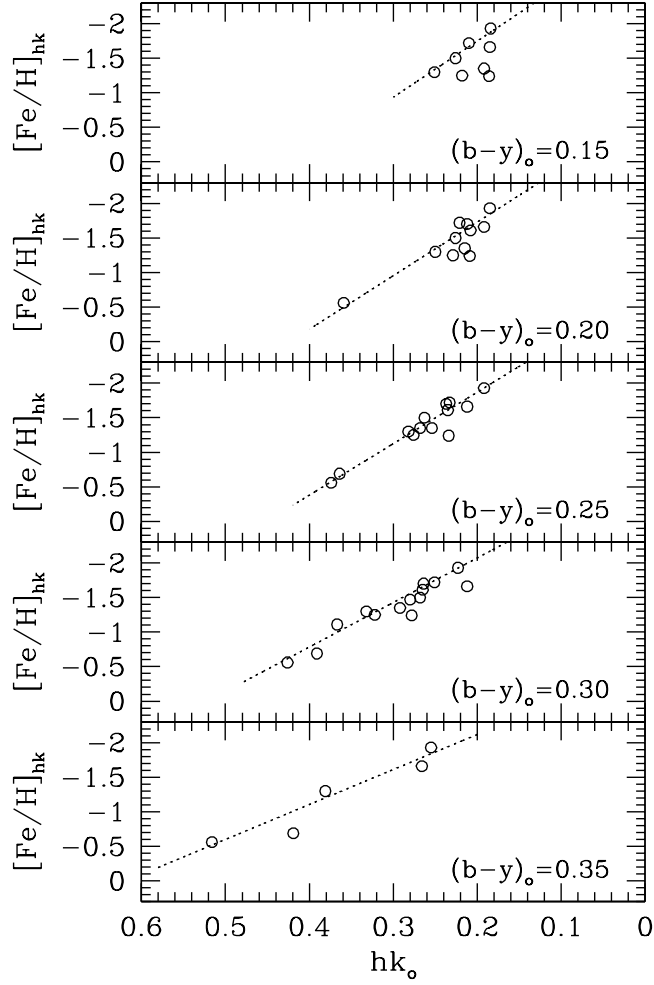}
 \caption{The relations between the metallicity measured from our $hk$ index, 
[Fe/H]$_{hk}$, and $hk_{o}$ at $(b-y)_{o}$ = 0.15, 0.20, 0.25, 0.30, and 0.35 
(Baird \& Anthony-Twarog 1999).
As temperature increases the sensitivity of $hk_{o}$ to [Fe/H]$_{hk}$ drops.}
\end{figure}

\begin{figure}
\centerline{\epsfysize=7in%
\epsffile{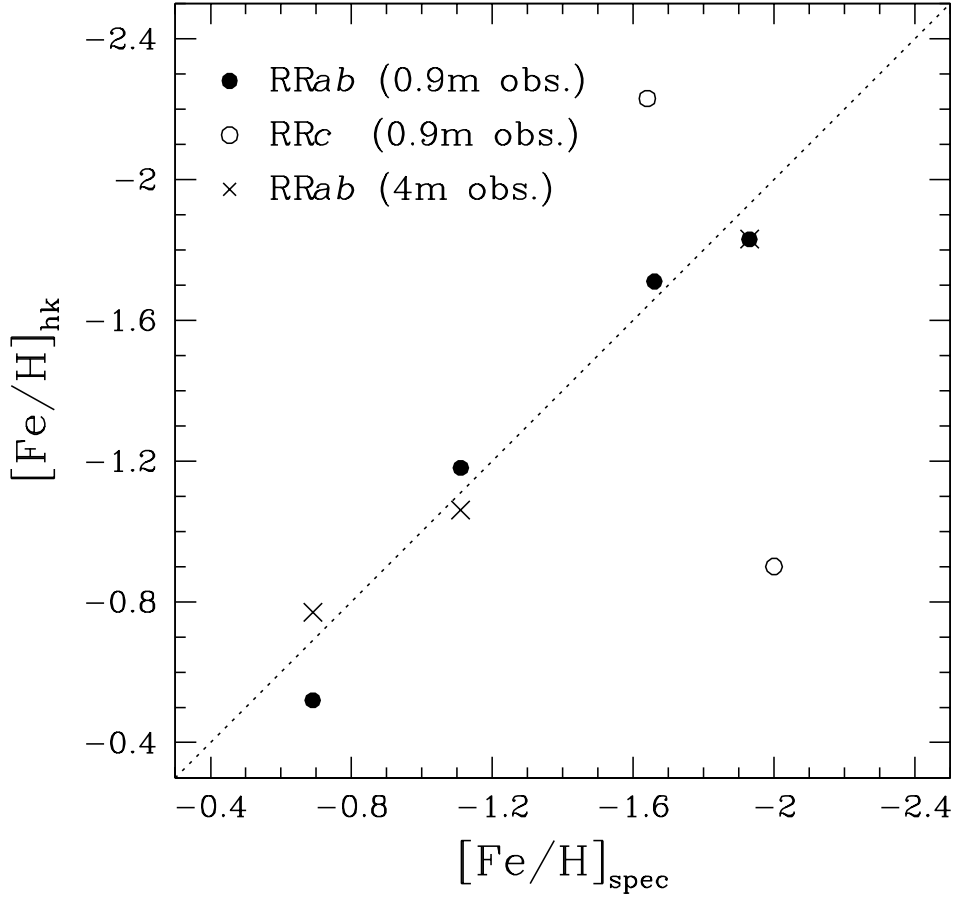}}
 \caption{Comparison between the metallicity measured spectroscopically, [Fe/H]$_{spec}$ 
(Baird 1996), and that from our $hk$ index, [Fe/H]$_{hk}$, for field RR Lyrae stars.
The [Fe/H]$_{hk}$ for the RR$ab$ stars are in excellent agreement with [Fe/H]$_{spec}$, 
but two RR$c$ stars show a larger scatter.}
\end{figure}

\begin{figure}
\centerline{\epsfysize=7in%
\epsffile{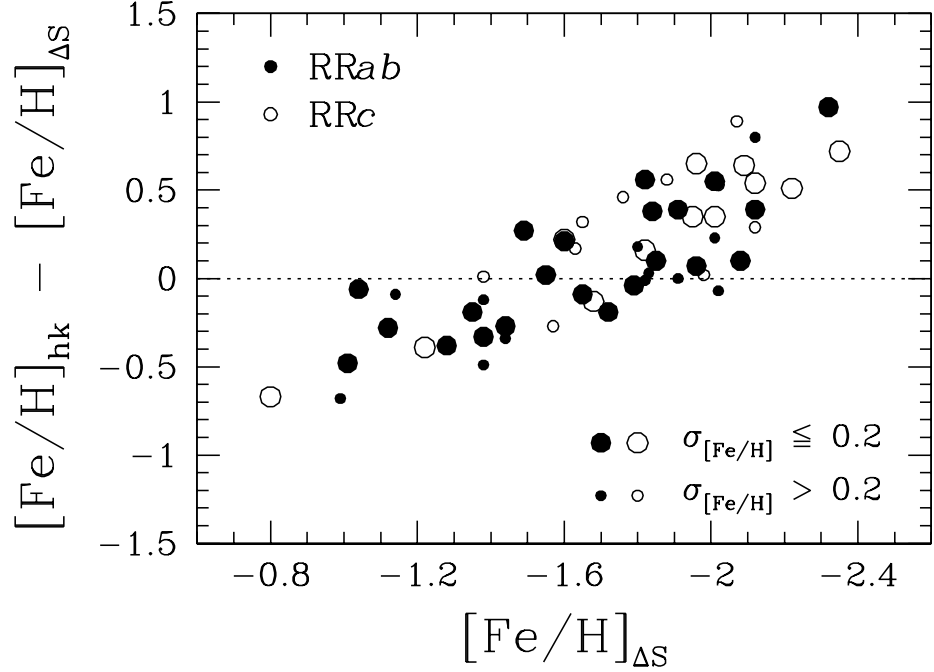}}
 \caption{Comparison between the metallicity measured from our $hk$ index, [Fe/H]$_{hk}$,
and that from previous $\Delta S$ observations, [Fe/H]$_{\Delta S}$, for 56 $\omega$ Cen 
RR Lyrae stars. The larger symbols are for stars with smaller ($\sigma_{[Fe/H]}$ $\leq$ 0.2 dex) 
observational error in [Fe/H]$_{hk}$. Note the significant differences between [Fe/H]$_{hk}$ and
[Fe/H]$_{\Delta S}$.}
\end{figure}

\begin{figure}
\centerline{\epsfysize=7in%
\epsffile{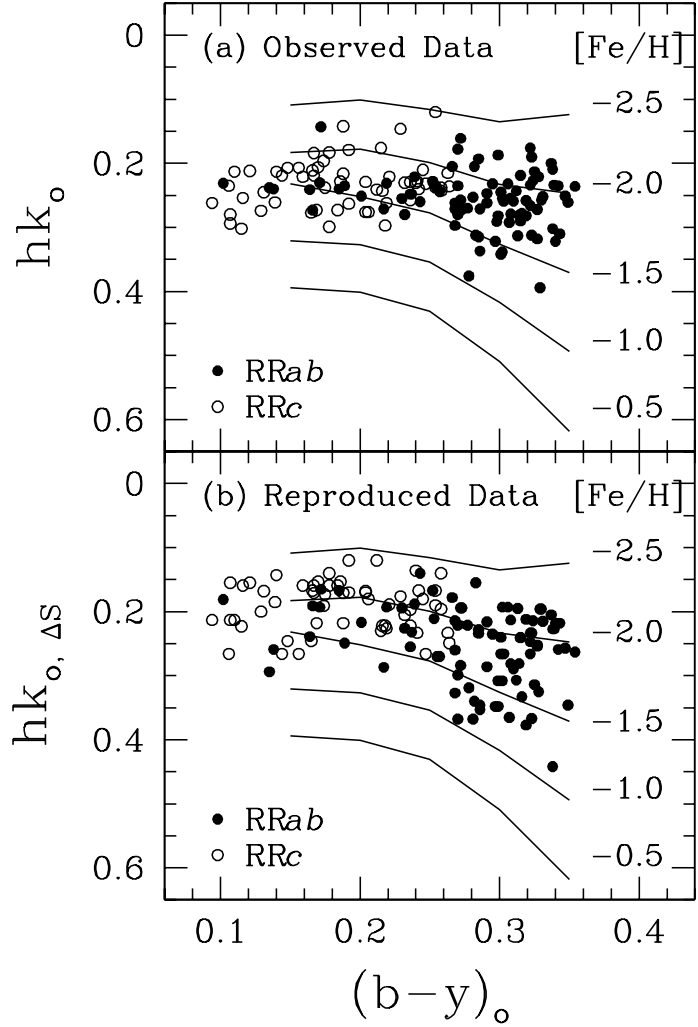}}
 \caption{Comparison between (a) observed $hk_{o}/(b-y)_{o}$ diagram and 
(b) reproduced $hk_{o, \Delta S}/(b-y)_{o}$ diagram.
Note that our observed $hk_{o}$ distribution of RR$ab$ stars is more
compressed than that of $hk_{o, \Delta S}$, which is the expected value from
the [Fe/H]$_{\Delta S}$, at any fixed $(b-y)_{o}$. 
In the case of most RR$c$ stars and some RR$ab$ stars with
$(b-y)_{o}$ $<$ 0.2, the distribution of $hk_{o}$ is shifted more metal rich
(about 0.5 dex) in the mean, compared to that of $hk_{o, \Delta S}$ (see text).}
\end{figure}

\begin{figure}
\centerline{\epsfysize=7in%
\epsffile{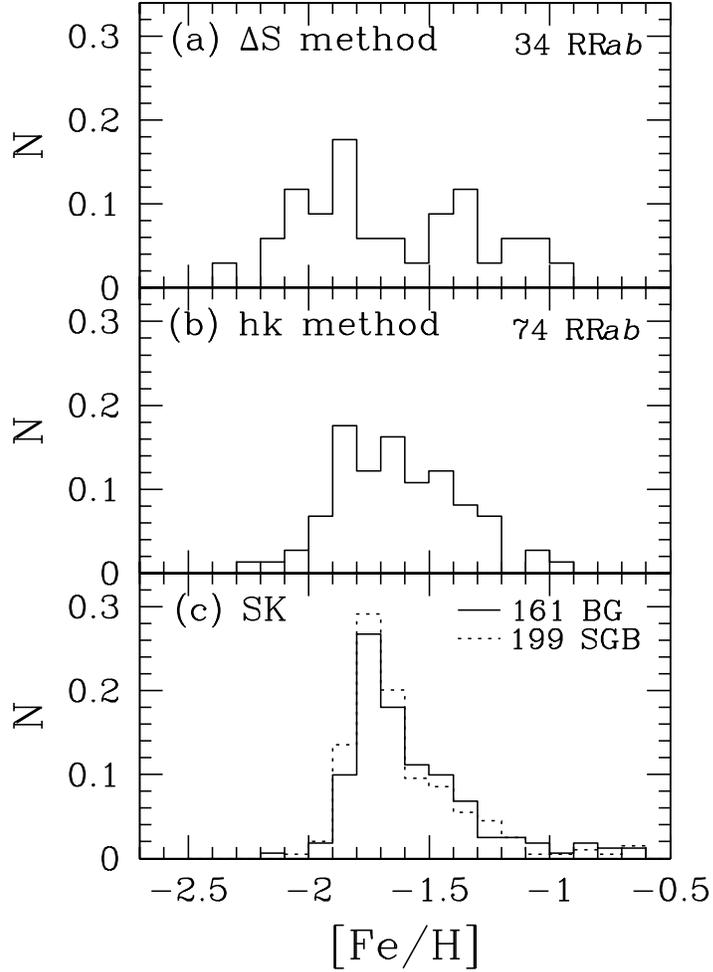}}
\caption{The metallicity distributions for (a) 34 RR$ab$ stars obtained from the
$\Delta$$S$ measurements, (b) our 74 RR$ab$ stars obtained from the $hk$ method,
and (c) 161 bright giants (BG; solid line) and 199 subgiants (SGB; dotted line)
from Suntzeff \& Kraft (1996). All metallicities are on the Zinn \& West (1984) scale.}
\end{figure}

\begin{figure}
\centerline{\epsfysize=6in%
\epsffile{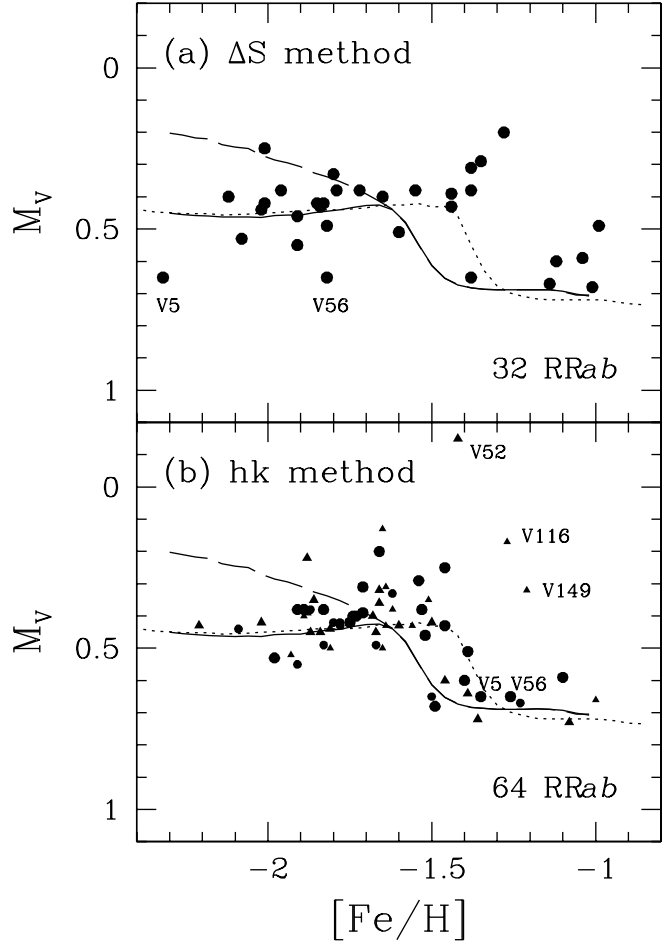}}
 \caption{The M$_{V}$(RR) - [Fe/H] relation of RR Lyrae stars in $\omega$ Cen
based on (a) the metallicity data for 32 RR$ab$ stars obtained from the $\Delta S$
method, and (b) the metallicity data for 64 RR$ab$ stars from our $hk$ method. 
The closed circles are stars overlapping with the 
sample of the $\Delta$$S$ measurements, while the closed triangles are stars 
available only from our photometry. The long-dashed line is a simple model
locus of Lee (1991) for the ensemble average of the RR Lyrae luminosities within
the instability strip with fixed mass loss, age, and $\alpha$-elements. The solid
(age = 13.5 Gyr) and short-dashed (age = 15.0 Gyr) lines are model loci which include
the nonmonotonic behavior of the horizontal-branch type with decreasing [Fe/H].
The M$_{V}$(RR) - [Fe/H]$_{hk}$ distribution appears to be in excellent agreement 
with the model loci (see text).} 
\end{figure}

\begin{figure}
\centerline{\epsfysize=7in%
\epsffile{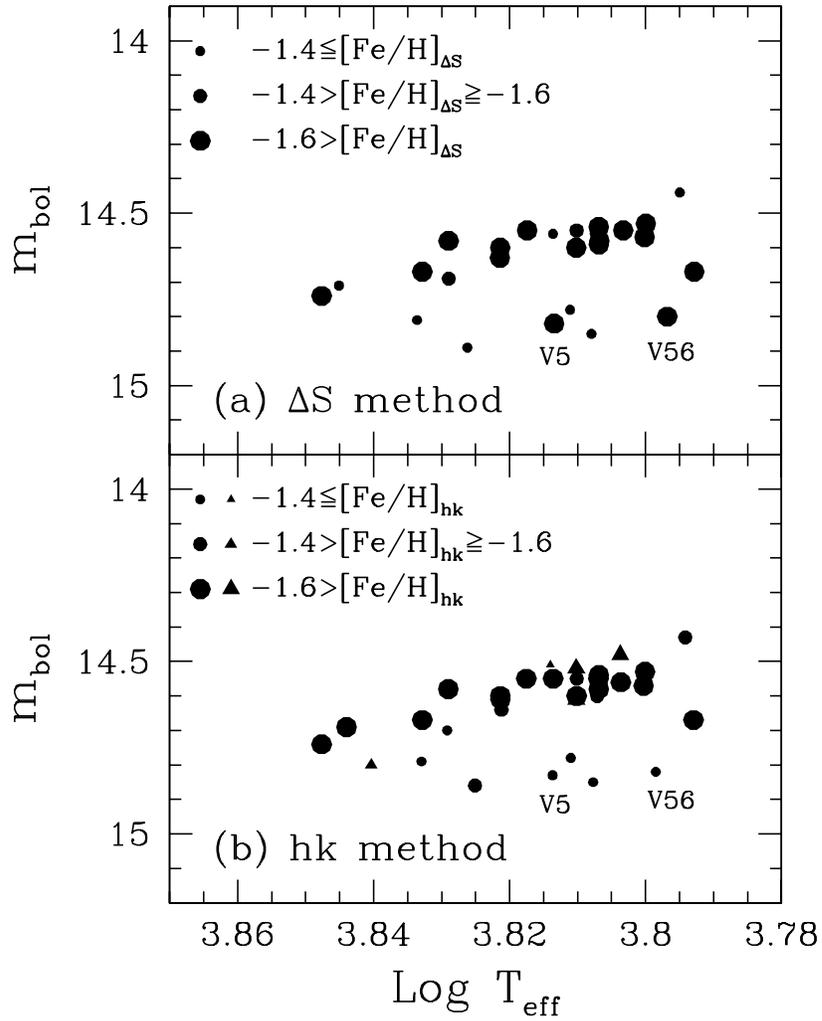}}
 \caption{(a) The m$_{bol}$ - logT$_{eff}$ diagram for 27 RR$ab$ stars in 
$\omega$ Cen, with metallicity data obtained from the $\Delta$$S$ method. There is
no clear relationship between metallicity and bolometric magnitude.
(b) Same as (a) for 34 RR$ab$ stars with metallicity data obtained from our
$hk$ method. The closed circles are stars overlapping with the sample of the 
$\Delta$$S$ measurement, while the closed triangles are variables only available from 
our photometry. Note that the metallicity dependence of m$_{bol}$ is more clearly 
defined in (b).}
\end{figure}

\begin{figure}
\centerline{\epsfysize=7in%
\epsffile{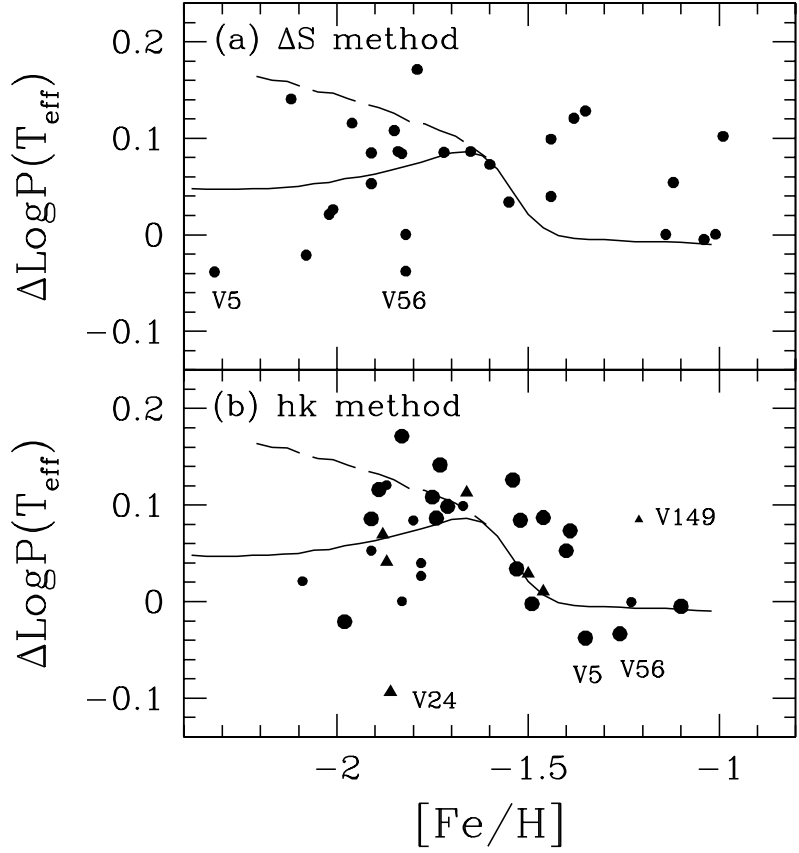}}
 \caption{(a) The $\Delta$log$P(T_{eff})$ - [Fe/H] diagram for 27 RR$ab$ stars 
in $\omega$ Cen, with metallicity data obtained from the $\Delta$$S$ method. 
As in Fig. 6, the solid and long dashed lines are model loci from Lee (1993).
(b) Same as (a) for 34 RR$ab$ stars with metallicity data obtained from our
$hk$ method. The closed triangles represent variables only available from
our photometry. Note again that our new observations agree better with the 
model predictions.}
\end{figure}

\begin{figure}
\centerline{\epsfysize=7in%
\epsffile{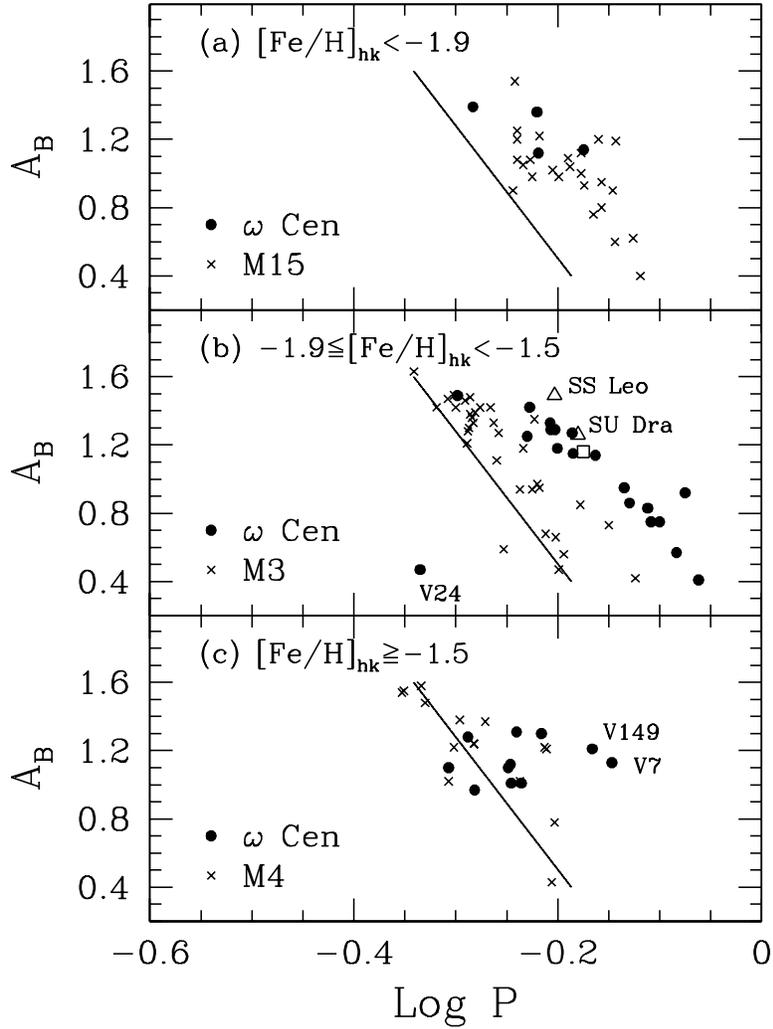}}
 \caption{(a) LogP-$A_{B}$ diagram for RR$ab$ stars with [Fe/H]$_{hk}$ $<$ -1.9.
The solid line is a fiducial line corresponding to the lower envelope 
of the M3 distribution, from
Sandage (1990a). (b) Same as (a) for stars with -1.9 $\le$ [Fe/H]$_{hk}$ $<$ -1.5.
We include two highly evolved field RR$ab$ 
stars, SU Dra and SS Leo (open triangles), from Jones et al. (1992). A highly evolved 
RR$ab$ star in M3, V65, is also marked with an open square (Clement \& Shelton 1999).
Most of $\omega$ Cen RR$ab$ stars in this metallicity range are believed to be 
highly evolved because they show an obvious deviation from M3 variables.
(c) Same as (a) for stars with [Fe/H]$_{hk}$ $\ge$ -1.5.} 
\end{figure}

\end{document}